\documentclass[aps,rmp,reprint,superscriptaddress,longbibliography,twocolumn]{revtex4-2}

\usepackage[table,cmyk]{xcolor}
\usepackage{amsmath, amsthm,commath,braket,booktabs}
\usepackage{calc,graphicx,bm,tikz}
\usepackage[charter,cal=cmcal,sfscaled=false]{mathdesign}

\usepackage{xr}
\usepackage{multirow}

\definecolor{oeawblue}{cmyk}{0.9,0.68,0,0}
\definecolor{iqoqiblue}{cmyk}{0.76,0.11,0,0}
\definecolor{iffsred}{cmyk}{0.12,0.94,0.87,0.34}
\definecolor{thpurple}{cmyk}{0.65,1.0,0.0,0.2}
\definecolor{uestcblue}{cmyk}{0.99,0.78,0.16,0.03}

\usepackage{hyperref}

\hypersetup{
  pdftitle={Variational Optimization for Quantum Problems using Deep Generative Networks},
  pdfauthor={Zhang, Lin et al.},
  pdfstartview=Fit,
  pdfpagelayout=SinglePage,
  colorlinks,
  linkcolor=uestcblue,
  citecolor=uestcblue,
  urlcolor=thpurple}
\usepackage[figure]{hypcap}

\usepackage[percent]{overpic}

\begin{document}

\title{Variational Optimization for Quantum Problems using Deep Generative Networks}

\author{Lingxia Zhang}\thanks{These authors contributed equally to this work}\affiliation{
Institute of Fundamental and Frontier Sciences and Ministry of Education Key Laboratory of Quantum Physics and Photonic Quantum Information,\\
University of Electronic Science and Technology of China, Chengdu 611731, China
}

\author{Xiaodie Lin}\thanks{These authors contributed equally to this work}\affiliation{
Institute for Interdisciplinary Information Sciences, Tsinghua University, Beijing 100084, China
}

\author{Peidong Wang}\affiliation{
Institute of Fundamental and Frontier Sciences and Ministry of Education Key Laboratory of Quantum Physics and Photonic Quantum Information,\\
University of Electronic Science and Technology of China, Chengdu 611731, China
}

\author{Kaiyan Yang}\affiliation{
Institute of Fundamental and Frontier Sciences and Ministry of Education Key Laboratory of Quantum Physics and Photonic Quantum Information,\\
University of Electronic Science and Technology of China, Chengdu 611731, China
}

\author{Xiao Zeng}\affiliation{
Institute of Fundamental and Frontier Sciences and Ministry of Education Key Laboratory of Quantum Physics and Photonic Quantum Information,\\
University of Electronic Science and Technology of China, Chengdu 611731, China
}

\author{Zhaohui Wei}\email{weizhaohui@mail.tsinghua.edu.cn}\affiliation{
Yau Mathematical Sciences Center, Tsinghua University, Beijing 100084, China
}
\affiliation{Yanqi Lake Beijing Institute of Mathematical Sciences and Applications, Beijing 101407, China}

\author{Zizhu Wang}\email{zizhu@uestc.edu.cn}\affiliation{
Institute of Fundamental and Frontier Sciences and Ministry of Education Key Laboratory of Quantum Physics and Photonic Quantum Information,\\
University of Electronic Science and Technology of China, Chengdu 611731, China
}

\begin{abstract}
Optimization drives advances in quantum science and machine learning, yet most generative models aim to mimic data rather than to discover optimal answers to challenging problems. Here we present a variational generative optimization network that learns to map simple random inputs into high quality solutions across a variety of quantum tasks. We demonstrate that the network rapidly identifies entangled
states exhibiting an optimal advantage in entanglement detection when allowing classical communication, attains the ground state energy of an eighteen spin model without encountering the barren plateau phenomenon that hampers standard hybrid algorithms, and—after a single training run—outputs multiple orthogonal ground states of degenerate quantum models. Because the method is model agnostic, parallelizable and runs on current classical hardware, it can accelerate future variational optimization problems in quantum information, quantum computing and beyond.
\end{abstract}

\maketitle
\tableofcontents


    \section{Introduction}
    Mathematical optimization is ubiquitous in modern science and technology. Spanning diverse fields like economics, chemistry, physics, and various engineering areas, its applications abound~\cite{kochenderferAlgorithmsOptimization2019}.  In quantum information theory, many problems relating to the approximation and characterization of quantum correlations can be formulated as convex optimization problems~\cite{PhysRevLett.98.010401,tavakoli2023-semidefinite,dohertyDistinguishingSeparableEntangled2002}, which is a particular kind of mathematical optimization with provable global optimality guarantees. For quantum problems where convexity is hard to come by or the global optimality of the solution is a secondary consideration when compared to the efficiency of the algorithm, variational optimization provides a rich toolbox. When solutions are expected to be quantum, hybrid quantum-classical variational algorithms are popular choices. In these algorithms, variational optimization is carried out on the classical parameters, while quantum gates and measurements are implemented in the corresponding quantum circuit~\cite{Peruzzo2014VQE, RevModPhys.92.015003,farhi2014quantum, PhysRevA.101.032308, Romero_2017, Havlicek:2019aa}.
    
    Optimization is also at the core of every machine learning algorithm~\cite{Murphy2022PML1}. Recently, machine learning algorithms have opened a new way to address scientific problems spanning a broad spectrum, accelerating the integration of AI into the scientific discovery process~\cite{Krenn2022AI4ScienceReview,Wang2023AI4ScienceReview}. In mathematics, they help humans discover new results~\cite{Davies2021AI4Math} and develop faster solutions to problems~\cite{DeepMindMatrixMul2022}. In biology, they help with drug developments~\cite{Jimenez-Luna2020AI4Med}. Particularly, generative models have seen explosive growth in the form of large language models~\cite{OpenAI2021GPT,Google2017Attention}, which are transforming the way humans interact with machines. Applying these models to science has enabled new solutions to mathematical problems to be discovered~\cite{FunSearch2023}. Meanwhile, generative models have also been widely applied to quantum physics. For example, many-body quantum models can be efficiently solved by restricted Boltzmann machines~\cite{carleoSolvingQuantumManybody2017, melkoRestrictedBoltzmannMachines2019}, lattice gauge theories can be simulated using normalizing flows~\cite{li2018-neurala, Stornati2022Variational}, quantum states can be more efficiently represented by variational autoencoders (VAEs)~\cite{kingma2013autoencoding,Rocchetto:2018aa, e21111091, carrasquillaReconstructingQuantumStates2019a}, and quantum circuits with desired properties can be generated by the generative pre-trained transformer~\cite{NakajiGQE24}.

    However, despite these encouraging advances, current applications of generative models to quantum problems usually focus on learning certain features from training data sets, and then generating new data with similar features. In the scenario where a classical (i.e., not quantum) generative model is used to solve a quantum problem, the training data may be quantum states or complex correlation information contained therein, and a neural network is expected to generate new quantum states or information resembling the training set. 
    
    In order to extend the possibility of applying generative models to quantum problems beyond this scenario, inspired by the classical variational autoencoder, we propose a method called the variational generative optimization network (VGON), whose output does not just resemble the input, but can be (nearly) optimal solutions to general variational optimization problems. VGON contains a pair of deep feed-forward neural networks connected by a stochastic latent layer, and a problem-specific objective function. The intrinsic randomness in the model can be leveraged both in its training and testing stages. During the training stage, we have not encountered any issues with the optimization getting trapped in local minima. We believe this can be partially explained by having random inputs, which effectively gives the optimization multiple starting points, and the architecture of our model, especially the existence of the latent layer, which regularizes the input and leads to good trainability. In the testing stage, the randomness allows VGON to produce multiple optimal solutions to the objective functions simultaneously, even after only a single stage of training. 

    We apply VGON to a variety of quantum problems to showcase its potential. We first demonstrate that it outperforms stochastic gradient descent (SGD) by avoiding entrapment in local optima in variational optimization problems of modest size, while also converging orders of magnitude faster. For larger problems with tens of thousands of parameters, we show that VGON can substantially alleviate the problem of barren plateaux in parameterized quantum circuits. Since generative models allow multiple optimal solutions to be found and generated simultaneously, a capability that deterministic algorithms lack, we use VGON to explore the ground state space of two quantum many-body models known to be degenerate. We show that VGON can successfully identify the dimensionality of the ground state space and generate a variety of orthogonal or linearly independent ground states spanning the entire space.

    \section{The VGON model}
\begin{figure}[!ht]
    \centering
    \includegraphics[width=0.98\columnwidth]{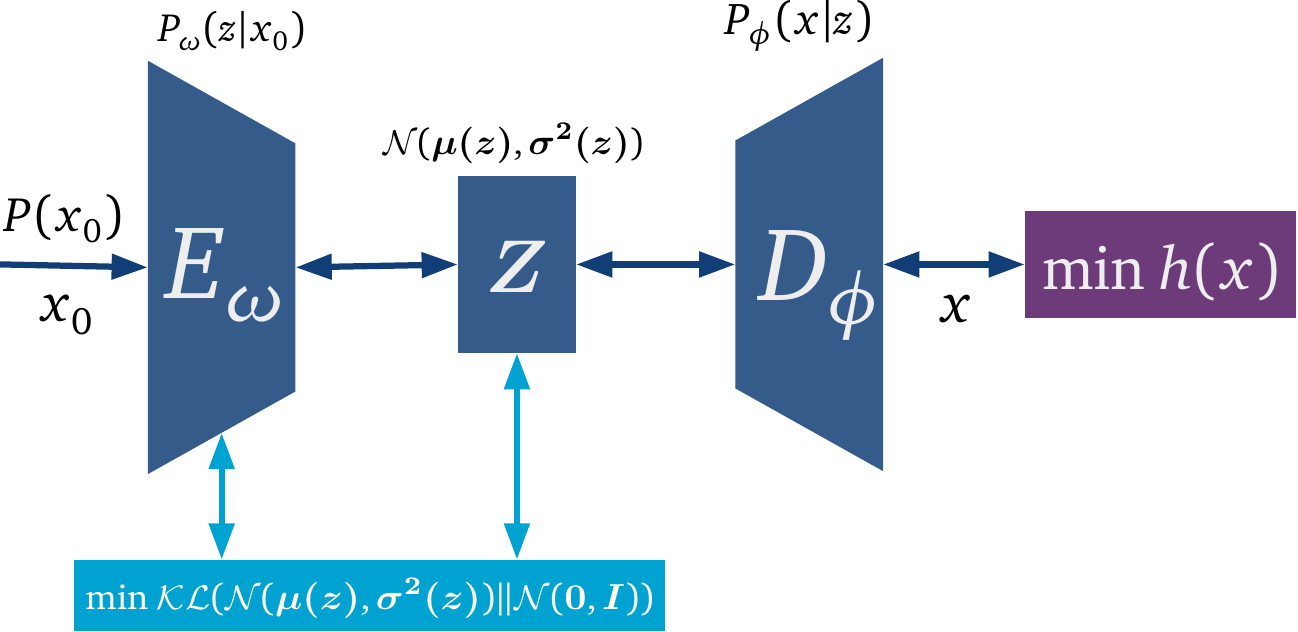}
    \caption{The framework of Variational Generative Optimization Network. The network is composed of an encoder network $E_{\bm{\omega}}$, a latent space $\mathcal{Z}$, and a decoder network $D_{\bm{\phi}}$. Training data $\bm{x_0}$ sampled from $P(\bm{x_0})$ is first mapped into a latent distribution $\mathcal{N}(\bm{\mu(z)},{\bm{\sigma^2(z)}})$ by $E_{\bm{\omega}}(\bm{x_0})$. Then a latent variable $\bm{z}$ sampled from $\mathcal{N}(\bm{\mu(z)},{\bm{\sigma^2(z)}})$ is transformed to the output $\bm{x}$ by $D_{\bm{\phi}}(\bm{z})$.
    The parameters $\bm{\phi}$ and $\bm{\omega}$ are updated iteratively to minimize the objective function $h(\bm{x})$, together with the Kullback-Leibler (KL) divervence between the latent distribution $\mathcal{N}(\bm{\mu(z)},{\bm{\sigma^2(z)}})$ and the standard normal distribution $\mathcal{N}(\bm{0},{\bm{I}})$.}\label{fig:vgon}
    \end{figure}

    The architecture of VGON, shown in Figure~\ref{fig:vgon}, consists of two deep feed-forward neural networks, the encoder $E_{\bm{\omega}}$ and the decoder $D_{\bm{\phi}}$ are connected via a latent layer $\mathcal{Z}$ containing a normal distribution $\mathcal{N}(\bm{\mu(z)},{\bm{\sigma^2(z)}})$, where the mean $\bm{\mu}$ and the standard deviation $\bm{\sigma}$ are provided by $E_{\bm{\omega}}$. During the training stage, input data $\bm{x_0}$ is sampled from a distribution $P(\bm{x_0})$, which in all our tests is the uniform distribution over the parameter space. It is then mapped to the latent distribution $\mathcal{N}(\bm{\mu(z)},{\bm{\sigma^2(z)}})$ by the encoder network $E_{\bm{\omega}}$. Next the decoder network $D_{\bm{\phi}}(\bm{z})$ maps data $\bm{z}$ sampled from the latent distribution $\mathcal{N}(\bm{\mu(z)},{\bm{\sigma^2(z)}})$ to a distribution minimizing the objective function $h(\bm{x})$. This minimization is achieved by iteratively updating  the parameters $\bm{\omega}$ and $\bm{\phi}$ in $E_{\bm{\omega}}$ and $D_{\bm{\phi}}$, respectively. Due to the existence of a stochastic latent layer, the gradients cannot be propagated backwards in the network. We solve this issue by using the reparameterization trick~\cite{kingma2013autoencoding}.
    
    The key difference between VGON and VAE lies in the objective function (also called the loss function in the machine learning literature): instead of asking the output data distribution to approximate the input distribution by maximizing a given similarity measure, VGON simply requires the output data to minimize any objective function that is appropriate for the target problem. In addition, the Kullback-Leibler (KL) divergence between the latent distribution $\mathcal{N}(\bm{\mu(z)},{\bm{\sigma^2(z)}})$ and the standard normal distribution $\mathcal{N}(\bm{0}, \bm{I})$ is also minimized during training, as part of the objective function. 

    After the objective function converges to within a given tolerance, the training stage is complete. To utilize the trained model, the encoder network is disabled, and data sampled from a standard normal distribution $\mathcal{N}(\bm{0},\bm{I})$ are fed into the decoder network. Depending on the characteristics of the objective function, the corresponding output distribution can be tightly centered around one or multiple optimum values. Moreover, it is worth mentioning that requiring the latent layer to follow a normal distribution not only facilitates efficient optimization of the objective function but also simplifies the sampling process, since the KL divergence between two normal distributions can be analytically evaluated and sampling from a normal distribution is computationally efficient.
    
    The goal of VGON can be seen as finding a way which maps a simple distribution defined over the latent space, i.e., the Gaussian, to a complicated one, i.e., a distribution whose samples can minimize the objective function with high probability~\cite{doerschTutorialVariationalAutoencoders2021}. This shares the spirit of transport theory, where given two probability measures $\mu$ and $\nu$ on spaces $X$ and $Y$, we call a map $T: X\to Y$ a transport map if $T_*(\mu)=\nu$, where $T_*(\mu)$ is the pushforward of $\mu$ by $T$, representing the process of transferring (or ``pushing forward'') the measure $\mu$ from $X$ to $Y$ via the measurable function $T$. In an optimal transport problem~\cite{villani2009optimal}, one is interested in finding a map $T$ that minimizes the transport cost $\int c(x, T_*(x))d\mu(x)$, subject to the constraint that the pushforward measure satisfies $T_*(\mu) = \nu$~\cite{peyre2019computational}.
    In addition to optimal transport problems, there are problems that do not have an explicitly defined target distribution, where the task is evaluating the loss function directly on the generated samples.
    In these problems, an optimal $T$ can either be found analytically, such as in inverse transform sampling, where both distributions are one-dimensional~\cite{devroyeNonUniformRandomVariate1986}, or $T$ can be learned/optimized from a parameterized $T_\theta$ on training data, such as generative model like normalizing flows~\cite{li2018-neurala,Wang2018MongeFlow}, where both distributions are high-dimensional but $T$ is invertible and constructed using neural networks. In VGON, the latent space usually has a much lower dimension than the output layer, making $T$ surjective, which means every point in the target space can be reached by applying the decoder to some latent input~\cite{nielsen2020survae}. This surjectivity relaxes the requirement for invertibility and enables VGON to easily cover the complex target distributions. In our experience, the best optimization results come from training the encoder-latent space-decoder triple as a whole, even though it is possible to achieve good results without including the encoder in the training process.

    To show that VGON can work well, we first use it to solve a variational optimization problem with a known unique optimal solution: finding the minimum ground state energy density among a class of quantum many-body models that matches the lower bound certified by an SDP relaxation~\cite{SDPreview24Mironowicz,SDPreview24Tavakoli}. More specifically, we consider a class of infinite 1D translation-invariant (TI) models with fixed couplings~\cite{Yang20221DContextuality}, and the optimization variables are the local observables. The ground state energy density of this class of models has a lower bound certified by a variant of the NPA hierarchy~\cite{Yang20221DContextuality}. However, there is no guarantee that any infinite TI quantum many-body Hamiltonian, if couplings are fixed but the local observables can be arbitrary, can achieve this bound. Meanwhile, by optimizing 3-dimensional local observables with SGD and computing the ground state with uniform matrix product state algorithms, a Hamiltonian whose ground state energy density matches the above lower bound to 7 significant digits has been found. We replace SGD with VGON to conduct the same optimization, and find that the converged model can (almost) deterministically generate Hamiltonians whose ground state energy density matches the NPA lower bound to 8 significant digits, reaching the precision limit of commercial SDP solvers (see Appendix~\ref{appendix:tdvp} for more technical details). Below we apply VGON to several more complicated problems.

     \section{Finding the optimal state for entanglement detection}
     Entanglement detection plays a central role in quantum tasks such as secure communication and distributed computing, where entanglement serves as a fundamental resource. Suppose two players, Alice and Bob, receive a bipartite quantum state $\rho$ from a source, then they want to determine whether $\rho$ is entangled, with the smallest statistical error.

    They can either perform the experiment independently in their respective laboratories and subsequently communicate the outcomes from Alice to Bob, or choose to forgo communication entirely. In the first scenario they are implementing a local operations and one-way classical communication (1-LOCC) protocol while in the second scenario they are implementing a local operations (LO) one. Implementing the 1-LOCC protocol experimentally requires fast real-time switching of Bob's measurement settings and a quantum memory to store Bob's half of the state while Alice performs her measurement and communicates the result. Do these extra experimental complexities yield tangible advantages such as reduced statistical error probabilities? In fact, it has been shown that for some simple states, such an advantage does exist, but it is too small to be useful~\cite{QuantumPrepareGame}. In fact, the advantage is highly dependent on the choice of target states and is hard to estimate analytically. The goal of this task is to identify high-dimensional quantum states for which this advantage, defined as the gap between the minimum statistical error probabilities in LO and 1-LOCC protocols, is as large as possible. This gap quantifies the practical advantage of allowing one-way communication between the parties in entanglement detection.

    Specifically, given a quantum state $\rho$, the advantage is defined to be the gap between the minimum probabilities $p_2$ of committing false-negative errors (a.k.a. type-II errors, defined as a source distributes an entangled state, but Alice and Bob conclude the state they received is separable) when Alice and Bob employ LO and 1-LOCC protocols, and the two protocols have the same probability of making false-positive errors, i.e., type-I error, denoted $p_1$ and defined as the scenario in which they conclude that they have received an entangled state, while the source actually distributes a separable one. The desired quantity can be computed by solving two SDPs that share the same objective but differ in constraint structure. Specifically, the following SDP is solved twice, once under LO protocol $P \in \mathcal{P}^{LO}$ and once under 1-LOCC protocol $P\in \mathcal{P}^{1\text{-}LOCC}$. The final result is obtained by subtracting the respective optimal values of $p_2$~\cite{Xing2025}:
    \begin{equation}
        \begin{aligned}\label{eq:sdp}
            \min_{P} \quad  & p_2 \\
            s.t.\quad & {\rm tr}(M_N(P)\rho)=p_2, \\
            & p_1 \mathbb{I}-M_Y(P) \in \mathcal{S}^*, \\
            & P \in \mathcal{P}^{\{LO,1-LOCC\}}.
        \end{aligned}
    \end{equation}
    Here $\mathcal{S}^*$ denotes the dual of the separable set $\mathcal{S}$. The positive operator-valued measure (POVM) operators $M_{Y (N)}(P)$ can be constructed as $ M_{Y (N)}(P) = \sum_{x,y, a,b}P(x,y, Y (N)|a,b)\;A^a_x\otimes B^b_y$, where $\{A_x^a\}$ ($\{B_y^b\}$) are predetermined measurements performed by Alice (Bob) with $x$ ($y$) being measurement labels and $a$ ($b$) being outcomes, and $P$ is the shorthand for the distribution $P(x,y, Y (N)|a,b)$, which specifies the detection strategy by assigning probabilities to particular combinations of settings, outcomes, and decisions. Here $Y$ and $N$ denote the decisions corresponding to the presence or absence of entanglement, respectively.

\begin{figure}[!ht]
    \centering
    \includegraphics[width=0.98\columnwidth]{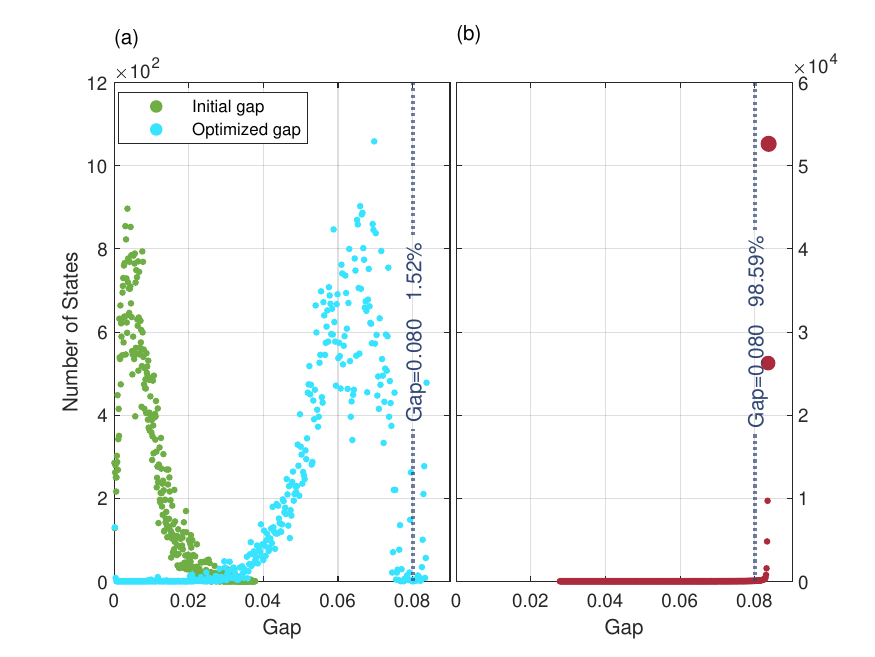}
    \caption{Comparisons of stochastic gradient descent (SGD) and Variational Generative Optimization Network (VGON) in generating states with large gaps. (a) Most of the 79,663 random initial states for SGD exhibit small gaps around $0.0036$, while after optimization $1.52\%$ of states have gaps larger than $0.08$, which is indicated by the dashed line. The largest gap is $0.083722$. (b) Over $98.59\%$ of the 100,000 states generated by a trained VGON model have gaps larger than $0.08$, which is presented by the dashed line. In particular, over $50\%$ of these states are tightly centered around $0.0837$.}
    \label{fig:gap}
    \end{figure}

    For a random quantum state $\rho$, it turns out that the gap calculated above is usually very small, as shown by the green dots in Figure~\ref{fig:gap}(a). In order to observe the gap under noisy experimental conditions, We focus on a linear optical setup that generates bipartite qutrit states, which also allows us to parameterize the state space. We first employ SGD to maximize the gap by starting from random pure bipartite qutrit states. The results are shown in Figure~\ref{fig:gap}(a). The SGD algorithm gets trapped easily in local maxima and needs to compute the gradient by solving dozens of SDPs. Optimizing the gap for 79,663 random pure states is computationally very costly (see Appendix~\ref{appendix:pure} for more details). Most of these states exhibit gaps around $0.0036$ before optimization, while the largest gap afterwards is $0.083722$. 
    
    \begin{table}[!ht]
    \centering
    \caption{Performance comparison between Variational Generative Optimization Network (VGON) and stochastic gradient descent (SGD). The comparison is carried out on finding states which maximize the advantage of one-way local operations and classical communication (1-LOCC) protocols over local operations (LO) ones.}
    
      \begin{tabular}{c  c  c  c}
      \toprule
      \multirow{2}{*}{\text{Method}}& \multicolumn{3}{c@{}}{\text{Percentage of states with gap}}\\
        \cmidrule(l){2-4}
       &  $\geq 0.08$ & $\geq 0.08355$ & $0.0837\pm 5\times 10^{-5}$\\
      \midrule
      {SGD} & \multirow{2}{*}{1.52\%} & \multirow{2}{*}{0.57\%} & \multirow{2}{*}{0.43\%}\\
      pure states & & &\\
      {VGON} & \multirow{2}{*}{98.59\%} & \multirow{2}{*}{52.75\%} & \multirow{2}{*}{9.38\%} \\  
      pure states & & &\\
      {VGON} & \multirow{2}{*}{99.32\%} & \multirow{2}{*}{57.36\%} & \multirow{2}{*}{22.19\%} \\
      mixed states & & &\\
      \bottomrule
    \end{tabular}
    \label{tbl:gap}
    \end{table}

    The results of using VGON to maximize the gap are depicted in Figure~\ref{fig:gap}(b) and summarized in Table~\ref{tbl:gap}. Based on 3,000 sets of initial parameters produced by uniform sampling, the model converges after less than two hours of training. After that, we use it to generate 10,000 output states. We find that over 98\% of them manifest gaps over 0.08, while over 50\% of them have gaps larger than $0.0835$. A similar performance has been observed even when choosing the initial quantum states from a variational submanifold of the space of all mixed states, where out of 10,000 states generated by the converged VGON model, 83 have gaps larger than $0.0837$, with an average purity of $0.99999$. For comparison, we also apply seven other global optimization algorithms and a Multilayer Perceptron (MLP)—a neural network with multiple fully connected layers—to this task, and find that VGON consistently outperforms all the baseline methods (see the second subsection in the Methods and Appendix~\ref{appendix:locc} for more details).
    Importantly, using VGON to solve an experiment-relevant instance of this problem allows us to experimentally demonstrate the advantage of 1-LOCC protocols over LO ones in detecting entanglement, where we can observe an error probability that is impossible to achieve with local operations alone~\cite{Xing2025}.

    \section{Alleviating the effect of barren plateaux in variational quantum algorithms}\label{sec:bp}
    
    On problems with a moderate size of optimization variables, VGON has shown its ability to quickly converge to the (nearly) optimal output distribution and generate high quality solutions with high probability. In near-term hybrid quantum-classical algorithms such as the variational quantum eigensolver (VQE)~\cite{Peruzzo2014VQE}, however, the number of classical parameters can quickly reach thousands or tens of thousands. The performance of such a hybrid algorithm can be hard to predict. On the classical part, the problems of vanishing gradients and having multiple minima are often present~\cite{NEURIPS2018_13f9896d,5264952,McCleanBarrenPlateau, Cerezo:2021aa, PRXQuantum.2.040316}. On the quantum part, the choice of ans\"{a}tze greatly affects the expressivity of the quantum circuit, making the certification of global optimality difficult~\cite{PhysRevResearch.3.023203,larocca2023-theoryb,Romero_2019,ucc,cerezo2024BP}.

    For example, in a typical VQE algorithm, a parameterized variational circuit $U(\bm{\theta})$ is used to approximately generate the ground state of a target Hamiltonian $H$. The circuit structure usually loosely follows the target Hamiltonian and is often called an ansatz. Then by setting the energy of the output state $|\psi(\bm{\theta})\rangle=U(\bm{\theta})|00\cdots0\rangle$ with respect to $H$, i.e., $\langle\psi(\bm{\theta})|H|\psi(\bm{\theta})\rangle$, as the objective function, the algorithm iteratively updates the parameters in the quantum circuit by applying gradient-based methods on a classical computer. When the algorithm converges, the output quantum state will likely be very close to the ground state of $H$.

    However, when the size of quantum systems increases, gradients vanish exponentially. This is primarily because the random initializations of parameterized unitaries conform to the statistics of a unitary 2-design~\cite{McCleanBarrenPlateau, Harrow:2009aa}, making the working of gradient-based optimization difficult.  To overcome the BP problem, several strategies have been proposed and investigated, with the small-angle initialization (VQE-SA) method being identified as an effective technique~\cite{PRXQuantum.3.010313, PRXQuantum.2.040309, PRXQuantum.3.020365}. It initializes parameters $\bm{\theta}$ to be close to the zero vector, which differs from the statistics of the parameters from a 2-design and thus may alleviate the BP problem. 
    
    The advantage of VGON over VQE-SA in alleviating BPs can be seen when we use them both, with the same parameterized quantum circuit, to compute the ground state energy of the Heisenberg XXZ model. Its Hamiltonian with periodic boundary conditions is given by
    \begin{align*}
        H_{XXZ}=-\sum_{i=1}^{N}\left(\sigma_x^{i}\sigma_{x}^{i+1}+\sigma_y^{i}\sigma_{y}^{i+1}-\sigma_z^{i}\sigma_{z}^{i+1}\right),
    \end{align*}
    where $\sigma_{x,y,z}^i$ denote the Pauli operators at site $i$.
    The ansatz for the parameterized quantum circuit is inspired by the matrix product state encoding~\cite{ran2020-encoding}. It consists of sequential blocks of nearest-neighbor unitary gates, each of which is made of 11 layers of single qubit rotations and CNOT gates (see Appendix~\ref{appendix:bp_xxz} for more details).

\begin{figure}[!ht]
    \centering
    \includegraphics[width=0.98\columnwidth]{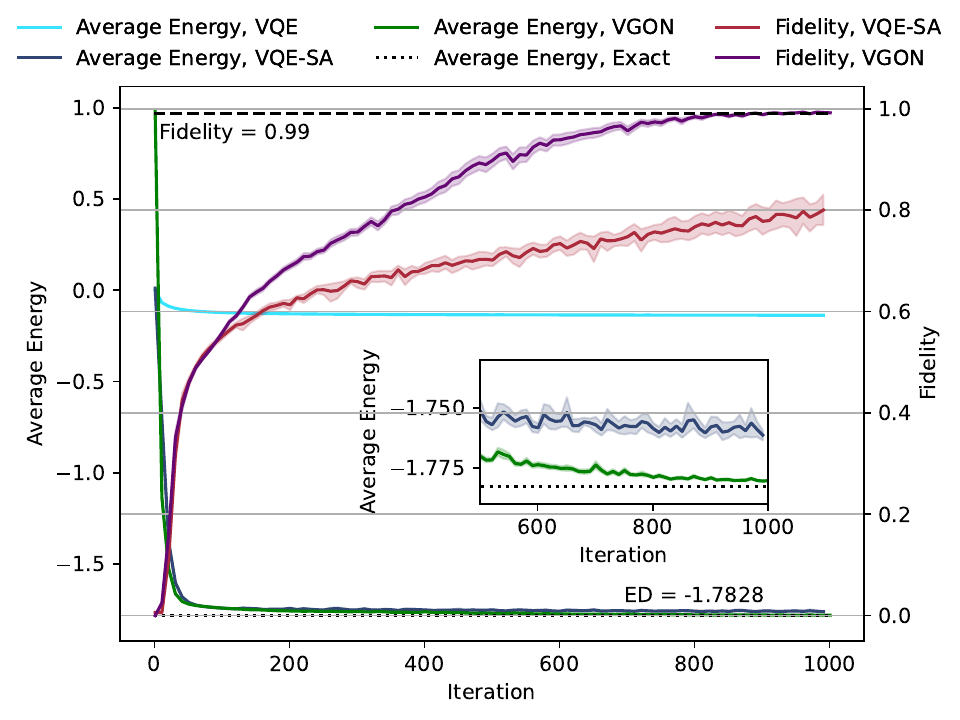}
    \caption{Mean values and 95\% confidence intervals of the energy densities and the fidelity to the exact ground state at different iterations. The light-blue line shows the average energy at different iterations for variational quantum eigensolver (VQE). The dark-blue (red) and the green (purple) lines represent the average energies (fidelity between the produced state and the exact ground state) at different iterations for the variational quantum eigensolver with small-angle initialization (VQE-SA) method and Variational Generative Optimization Network (VGON), respectively. The exact average ground state energy is depicted by the black dots. The inset zooms in on the convergence behavior of the average energies for VGON and VQE-SA, showcasing the faster convergence of VGON. Each method is repeated 10 times to calculate the mean values and 95\% confidence intervals.}
    \label{fig:res_BP}
    \end{figure}

    By choosing $N=18$, the circuit {contains $816$ blocks} and $12,240$ variational parameters. The average ground state energy, computed using exact diagonalization (ED), is -1.7828. We use both VQE-SA and VGON to compute the same quantity, with each method repeated 10 times. The results are shown in Table~\ref{tbl:BP} and Figure~\ref{fig:res_BP}, where the mean values and the 95\% confidence intervals of these methods are visualized. The dark-blue and the green lines represent the average energy for VQE-SA and VGON, whose mean values at the last iteration are -1.7613 and -1.7802, respectively. Furthermore, to compare the performance of the two methods in a more fine-grained manner, we also calculate the fidelity between the states produced by the quantum circuit and the exact ground state. As the purple line depicted, VGON can achieve a 99\% fidelity by around 880 iterations, while the VQE-SA method can only achieve 78.25\% fidelity within the same number of iterations. 
    We would like to remark that since the quantum computational resource consumed in VGON is similar to that in VQE-SA and classical computational resource is relatively cheap, the wall clock time cost per iteration for VGON can be comparable to that of VQE-SA. Moreover, the batch training of VGON can lead to a more stable convergence. For practical on-hardware quantum optimization, batch evaluation also enhances noise robustness by averaging over multiple circuit executions, thereby mitigating stochastic fluctuations, suppressing outliers, and accelerating convergence. For another comparison, VQE with uniformly random initial parameters can barely provide meaningful results due to the presence of barren plateaux, where the mean value of the average energy is -0.1367 after 1,000 iterations across 10 repetitions, as illustrated by the light-blue line in Figure~\ref{fig:res_BP}. 

    \begin{table*}[!ht]
      \centering
      \caption{Comparison between variational quantum eigensolver (VQE), VQE with small-angle initialization (VQE-SA), and Variational Generative Optimization Network (VGON).}
      \fontsize{8}{10}\selectfont
      \setlength{\tabcolsep}{6pt}
      \begin{tabular}{c c c c c c c}
      \toprule
      \multirow{2}{*}{\text{-}}& \multicolumn{3}{c@{}}{\text{Optimal}} & \multicolumn{3}{c@{}}{\text{Mean}}\\
        \cmidrule(l){2-7}
       &  VQE & VQE-SA & VGON & VQE & VQE-SA & VGON\\
      \midrule
      Average Energy & -0.1374 & -1.7684 & -1.7821 & -0.01367 & -1.7613 & -1.7802\\[6pt]
      Fidelity & - & 90.00\% & 99.82\% & - & 80.01\% & 99.17\% \\[3pt]
      \toprule
    \end{tabular}
    \label{tbl:BP}
    \end{table*}
    
    \section{Identifying degenerate ground state space of quantum models}
    Deterministic gradient-based optimization methods are predisposed to follow a single path, therefore hampering their ability to efficiently detect multiple optima. A unique advantage of generative models is the ability to produce diverse samples of objects, all of which may minimize the objective function. In optimization, this leads to the possibility of finding multiple optimal solutions with a single stage of training. We now show that when appropriately trained, VGON exemplifies such an effective capability for generating multiple (nearly) optimal solutions simultaneously. This capability can be largely ascribed to its integration of randomness and the adoption of batch training. The former facilitates broader exploration within the variational manifold, while the latter, which involves processing subsets of data samples concurrently, supports the collective identification of multiple optimal solutions. 

    A natural multi-optima problem in quantum many-body physics is the exploration of degenerate ground spaces of quantum many-body Hamiltonians. We apply VGON to two Hamiltonians with known degenerate ground states: the Majumdar-Ghosh (MG)~\cite{majumdarNextNearestNeighbor1969,majumdarNextNearestNeighborInteractionLinear1969} model in {Eq.~\eqref{eq:degen_mg}}, and a Heisenberg-like model {in Eq.~\eqref{eq:degen_232}} coming from one of the contextuality witnesses presented in {Ref.}~\cite{Yang20221DContextuality}:
    \begin{align}
     H_{MG} &= \sum_{i=1}^{N} \bm{\sigma}^i\cdot \bm{\sigma}^{i+1} + \bm{\sigma}^{i+1} \cdot\bm{\sigma}^{i+2}+ \bm{\sigma}^i \cdot\bm{\sigma}^{i+2},\label{eq:degen_mg}\\
     H_{232}&=\sum_{i=1}^{N}(2\sigma_{x}^i\sigma_{x}^{i+1}+\sigma_{x}^i\sigma_{y}^{i+1}-\sigma_{y}^i\sigma_{x}^{i+1}),\label{eq:degen_232}
    \end{align}
    where $\bm{\sigma}^i = (\sigma^i_x, \sigma^i_y, \sigma^i_z)$ are Pauli operators at site $i$. We take $N=10$ for $H_{MG}$, and $N=11$
    for $H_{232}$, making their ground state spaces 5- and 2-fold degenerate, respectively. An orthonormal basis for their respective degenerate ground state spaces is computed by the ED method, which outputs five vectors $\ket{v_1}\ldots\ket{v_5}$ spanning the ground state space of $H_{MG}$, and two vectors $\ket{u_1}$ and $\ket{u_2}$ spanning that of $H_{232}$. 

    The overall objective of this problem is similar to the previous one: finding the ground states of $H_{MG}$ and $H_{232}$ with variational quantum circuits. We maintain the same circuit layout as in the previous problem, and use 36 and 60 {blocks of unitary gates} for each Hamiltonian respectively. Profiting from the use of mini-batches to estimate gradients, a common technique in training neural networks, VGON can effectively evaluate many different circuits simultaneously. Meanwhile, to enhance intra-batch diversity, a penalty term consisting of the mean cosine similarity among all pairs of sets of circuit parameters in the same batch is added to the objective function. This penalty term, together with the mean energy of the states in the batch, ensures a balance between maintaining the diversity of the generated outputs and minimizing the energy. Further details can be found in Appendix~\ref{appedix:bp_grads}.
    
    As a result, unlike VQE-based algorithms aiming to generate multiple energy eigenstates, the objective function of VGON is model-agnostic. In other words, with no prior knowledge of the degeneracy of the ground space, VGON is capable of generating orthogonal or linearly independent states in it. In comparison, to achieve a diversity of outputs with VQE-based algorithms~\cite{Higgott2019variationalquantum,PhysRevResearch.1.033062}, it is essential to provide diverse inputs for the VQE model. However, attaining this diversity can result in barren plateaux within the optimization landscape. Though employing VQE-SA may address this problem, it could significantly diminish the diversity of inputs, as it tends to constrain inputs to values near zero. 

\begin{figure}[!htbp]
    \centering
    \includegraphics[width=0.98\columnwidth]{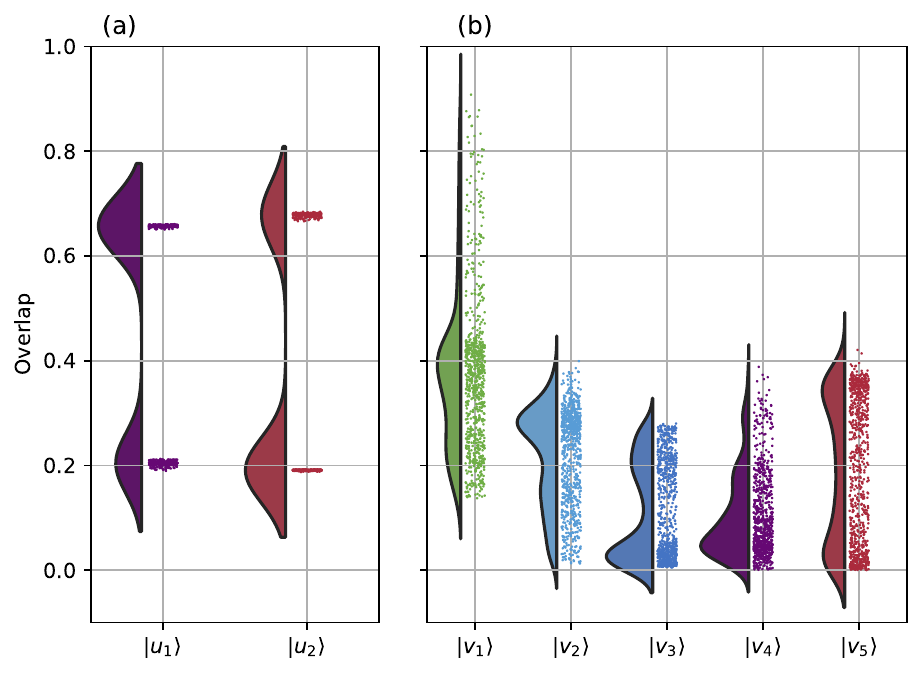}
    \caption{The overlap between 1,000 states generated by the trained Variational Generative Optimization Network and the orthonormal bases of the ground space. The corresponding orthonormal bases of the ground space are computed by exact diagonalization, with notations $\ket{u_1}$ and $\ket{u_2}$ for $H_{232}$ shown in (a) and $|v_{1}\rangle,|v_{2}\rangle,\cdots,|v_{5}\rangle$ for $H_{MG}$ shown in (b). The shaded curves show the population densities of the generated states having different overlaps with one of the basis states.}
    \label{fig:degen_rain}
    \end{figure}

    We generate 1,000 output states for each Hamiltonian using a VGON model trained with the above objective function. We find that the vast majority of these states have energy low enough to be treated as ground states. Figure~\ref{fig:degen_rain} shows the overlap between the generated states and the basis of their ground state space. In Figure~\ref{fig:degen_rain}(a), the generated states for $H_{232}$ fall into two orthogonal classes, which form an orthonormal basis of the ground state space. For $H_{MG}$, Figure~\ref{fig:degen_rain}(b) shows that most of them are linearly independent and span the same space as $\ket{v_1}\ldots\ket{v_5}$.

\section{Details}
\subsection{Details on VGON}
A VGON model contains two neural networks, i.e., the encoder $E_{\bm{\omega}}:\mathcal{X}\rightarrow\mathcal{Z}$ and the decoder $D_{\bm{\phi}}:\mathcal{Z} \rightarrow \mathcal{X}$ connected by a latent space $\mathcal{Z}$, and they are parameterized by $\bm{\omega,\phi}$, respectively. These parameters are iteratively updated to produce a solution distribution $P_{\bm{\phi,\omega}}(\bm{x})$ such that the expectation of an objective function $h(\bm{x})$ is optimized:
\begin{align*}
   &\mathbb{E}_{\bm{x}\sim P_{\bm{\phi,\omega}}(\bm{x})}\left[h(\bm{x})\right]\\
    =&\int \int h\left(D_{\bm{\phi}}(\bm{z})\right)P_{\bm{\omega}}(\bm{z}|\bm{x_0})P(\bm{x_0})d\bm{x_0}d\bm{z}\\
    =&\int h\left(D_{\bm{\phi}}(\bm{z})\right)P_{\bm{\omega}}(\bm{z})d\bm{z}\\
    =&\mathbb{E}_{z \sim P_{\bm{\omega}}(\bm{z})} \left[h\left(D_{\bm{\phi}}(\bm{z})\right)\right].
\end{align*}
More specifically, the input data $\bm{x_0}$ is sampled from a given distribution $P(\bm{x_0})$, which is then mapped to the latent distribution $P_{\bm{\omega}}(\bm{z})$ by the encoding estimator $E_{\bm{\omega}}$, i.e., $P_{\bm{\omega}}(\bm{z})=\int P_{\bm{\omega}}(\bm{z}|\bm{x_0})P(\bm{x_0})d\bm{x_0}$. Next the decoding estimator $D_{\bm{\phi}}(\bm{z})$ further maps the latent distribution $P_{\bm{\omega}}(\bm{z})$ to a distribution $P_{\bm{\phi,\omega}}(\bm{x})$ of the target data $\bm{x}$, which is right the input of the objective function $h(\bm{x})$.

Additionally, for the convenience of training, we constraint the distribution of the latent space to be a normal distribution $\mathcal{N}(\bm{\mu(z)},{\bm{\sigma^2(z)}})$, and try to minimize the distance between it and the standard normal distribution $\mathcal{N}(0,I)$, measured by the KL divergence. Specifically, the cost function for VGON is formulated as

\begin{equation}\label{eq:loss_func}
\begin{aligned}
&C(\bm{\phi,\omega})\\
=&\mathbb{E}_{\bm{x}\sim P_{\bm{\phi,\omega}}(\bm{x})}\left[h(\bm{x})\right]
+\beta\cdot \mathcal{D}\left[\mathcal{N}(\bm{\mu(z)},{\bm{\sigma^2(z)}})||\mathcal{N}(0,I)\right],
\end{aligned}
\end{equation}
where the hyperparameter $\beta$ represents the trade-off between the expectation of the objective function and the above KL divergence. 

In our implementations of VGON, all the training procedures are conducted based on the PyTorch framework~\cite{NEURIPS2019_bdbca288}. To address different tasks, diverse objective functions $h(\bm{x})$ are employed, and each requires a specific interfacing with PyTorch. The configurations of these VGON models will be detailed in the subsequent sections, which can provide us a comprehensive understanding of how VGON is tailored for different optimization challenges.

\subsection{Datails on finding the optimal states in entanglement detection}\label{method_loss}

To find the quantum state that can exhibit the largest advantage of 1-LOCC protocols over LO protocols in entanglement detection, the problem can be formulated as maximizing the difference between the solutions to the two SDPs introduced in Eq.~\eqref{eq:sdp}, with the following objective function and parameter space:
\begin{itemize}
		\item \textbf{Objective function}: $h(\rho',p_1)={p_2^{LO}}^*(\rho',p_1)-{p_2^{1-LOCC}}^*(\rho',p_1)$
		\item \textbf{Parameter space}: $\{e_1\in \mathbb{R}, \rho'\in \{\rho_{exp}\} \text{ or } \{\rho\}\}$
\end{itemize}
Here, $p_1$ is parameterized as $p_1=({\rm tanh}(e_1)+1)/2$, and $\{\rho_{exp}\}$ and $\{\rho\}$ represent the set of states for the pure case and mixed case, respectively.

Efficient variational optimization for these SDPs and their integration into the PyTorch framework for machine learning requires the use of CVXPY and cvxpylayers~\cite{Diamond2016cvxpy,Agrawal2018rewrite,Agrawal2019cvxpylayers}. The first translates a convex optimization problem into a form that solvers can understand, while the latter allows automatic differentiation of convex optimization problems by computing their gradients and backpropagate them through the neural network.

As we mentioned in the main text, the state space we consider follows closely the linear optical setup which generates arbitrary bipartite qutrit states $\rho_{\rm exp}$. Photons generated by the laser source are expressed as $\sum_i c_i|i \rangle$, where $c_i$ are complex numbers satisfying $\sum_i |c_i|^2=1$. Afterwards, these photons go through spontaneous parametric down-conversion (SDPC), which converts their state to $|\psi \rangle=\sum_i c_i|ii \rangle$. In the case of qutrits, the state can be parameterized as~\cite{OptimalDec,Hu2018}
\begin{align*}
    |\psi \rangle = \sin \frac{\theta}{2} \cos \frac{\phi}{4} e^{im} |00 \rangle +
    \sin \frac{\theta}{2} \sin \frac{\phi}{4} e^{in} |11 \rangle+
    \cos \frac{\theta}{2}|22 \rangle,
\end{align*}
where $\phi,m,n \in [0,2\pi)$, and $\theta \in [0,\pi]$ are variational parameters.
Subsequently, two local unitaries denoted by $U_A, U_B$ can be applied on the two subsystems, resulting in the quantum state
\begin{align*}
\rho_{\rm exp}=(U_A\otimes U_B)|\psi\rangle\langle\psi|(U_A^{\dagger}\otimes U_B^{\dagger}).
\end{align*}
Here, $U_A$ ($U_B$) can be parameterized by a set of $3^2=9$ linearly-independent skew-Hermitian matrices $\{T_j\}$ \cite{hylandLearningUnitaryOperators2017}, i.e.,
\begin{align*}
    U_A\ (U_B) = \exp\left(\sum_{j = 1}^{9}\lambda_j T_j\right),
\end{align*}
where $\lambda_j$'s are $9$ real numbers, denoted as $\bm{\lambda}_A$ ($\bm{\lambda}_B$). Therefore, the parameterized space for pure states $\rho_{\rm exp}$ is represented as
\begin{align*}
    \{m,n,\phi,\theta \in \mathbb{R},\ \bm{\lambda}_A,\bm{\lambda}_B\in\mathbb{R}^{9}\}.
\end{align*}
On the other hand, mixed qutrit states $\rho\in\mathcal{H}^3\otimes\mathcal{H}^3$ can be parameterized by 
\begin{align*}
    \rho=U\Sigma U^{\dagger},
\end{align*}
where $\Sigma$ is a $9\times 9$ diagonal matrix whose diagonal entries are nonnegative and sum to $1$, and $U$ is a unitary matrix that can be parameterized by a set of $9^2-1$ generalized Gell-Mann matrices $\{T_j\}$~\cite{Bertlmann2008}, i.e.,
\begin{align*}
    U=\exp\left(i\sum_{j=1}^{9^2-1}\lambda_jT_j\right),
\end{align*}
where $\lambda_j$'s are $9^2-1$ real numbers, denoted as $\bm{\lambda}\in\mathbb{R}^{9^2-1}$. Furthermore, the normalized diagonal matrix $\Sigma$, denoted as ${\rm diag}(\sigma_1^2,\cdots,\sigma_9^2)$, can be obtained by ensuring that the Euclidean norm of the vector $\bm{\sigma}$ is equal to 1, i.e., $|\bm{\sigma}|_2=1$, where $\bm{\sigma}=(\sigma_1,\cdots,\sigma_9)$. Consequently, the parameterized space for mixed states $\rho$ case is written as
\begin{align*}
    \{\bm{\lambda}\in\mathbb{R}^{9^2-1}, \bm{\sigma}\in\mathbb{R}^9:|\bm{\sigma}|_2=1\}.
\end{align*}

\subsection{Details on alleviating barren plateaus in variational quantum algorithms}

A typical VQE algorithm can approximate the ground state of a given Hamiltonian $H$ using a variational wave function generated by a parameterized quantum circuit (PQC) $U(\bm{\theta})$, represented as $|\psi(\bm{\theta})\rangle=U(\bm{\theta})|00\cdots0\rangle$. This sets up the minimization problem:
\begin{itemize}
    \item \textbf{Objective function}: $h(\bm{\theta})=\langle\psi(\bm{\theta})|H|\psi(\bm{\theta})\rangle$
    \item \textbf{Parameter space}: $\{\bm{\theta}\in\mathbb{R}^{M}\}$
\end{itemize}
The dimension of the parameter space $M$ is determined by the structure of the PQC. 

The simulation of PQCs and the computation of energy are implemented by PennyLane~\cite{bergholm2022pennylane}, a software library for quantum machine learning. Its support of the CUDA-based CuQuantum SDK from NVIDIA enables VGON to handle over 10000 variational parameters on a consumer grade graphics card. PennyLane also provides seamless integration with PyTorch and its machine learning toolkit. Efficient GPU-accelerated simulation of PQCs is achieved by using the adjoint differentiation method~\cite{jones2020efficient} to compute the gradients, after which the parameters are updated by the Adam optimizer. 

One of the key differences between VQE, VQE-SA and VGON is the initialization of parameters. For the VQE and VQE-SA, the initial parameters $\bm{\theta}$ are uniformly sampled from the range $[0,2\pi)$ and $[0,0.01)$, respectively. In VGON, on the other hand, the decoder initialized using PyTorch's default settings generates the circuit's initial parameters $\bm{\theta}$. For more details on these methodologies and the comparisons between their performance, please refer to the {third subsection in the Results} and Appendix~\ref{appendix:locc}.

\subsection{Details on identifying degenerate ground state space of quantum models}
To identify the degenerate ground space of a Hamiltonian $H$ with VGON, the objective function needs two pivotal components to steer the optimized quantum state $|\psi(\bm{\theta})\rangle$ towards diverse ground states. The first component utilizes a PQC $U(\bm{\theta})$ to generate the state $|\psi(\bm{\theta})\rangle=U(\bm{\theta})|00\cdots 0\rangle$, targeting the ground space. The second component integrates a cosine similarity measure into the optimization objective, aiming to enhance the diversity among the generated quantum states.

Specifically, for a batch of $S_b$ states $\{|\psi(\bm{\theta}_i)\rangle\}$, the mean energy is calculated by
\begin{align*}
    \bar{E}(\bm{\Theta})=\frac{1}{S_b}\sum_{i=1}^{S_b}\langle\psi(\bm{\theta}_i)|H|\psi(\bm{\theta}_i)\rangle,
\end{align*}
where $\bm{\Theta}=(\bm{\theta}_1, \bm{\theta}_2, \cdots, \bm{\theta}_{S_b})$. In addition, a penalty term for the objective function based on the cosine similarity is defined as
\begin{align*}
    \bar{S}_{\mathcal{C}_{S_b}^2}(\bm{\Theta})=\frac{1}{|\mathcal{C}_{S_b}^2|}\sum_{(i,j)\in \mathcal{C}_{S_b}^2}\frac{\bm{\theta}_i\cdot\bm{\theta}_j}{\Vert\bm{\theta}_i\Vert\Vert\bm{\theta}_j\Vert},
\end{align*}
where $\mathcal{C}_{S_b}^2$ represents the set of all 2-combinations pairs derived from the elements in $\{1,2,\cdots,S_b\}$, and $\Vert\cdot\Vert$ denotes the Euclidean norm. Eventually, the optimization objective is set as minimizing the linear combination of $\bar{E}(\bm{\Theta})$ and $\bar{S}_{\mathcal{C}_{S_b}^2}(\bm{\Theta})$ according to a trade-off coefficient $\gamma$, i.e.,
\begin{itemize}
\item \textbf{Objective function}: $h(\bm{\Theta}) = \bar{E}(\bm{\Theta})+\gamma \cdot \bar{S}_{\mathcal{C}_{S_b}^2}(\bm{\Theta})$
\item \textbf{Parameter space}: $\{\bm{\Theta}\in\mathbb{R}^{S_bM}\}$
\end{itemize}

We estimate the quality of the generated state by computing the overlap between them and a basis of the ground space. Such computations can be resource-intensive, and therefore we only demonstrate the performance of VGON for 10-qubit systems.

\section{Conclusions}
We propose a general approach called variational generative optimization network, or VGON for short, for tackling variational optimization challenges in a variety of quantum problems. This approach combines deep generative models in classical machine learning with sampling procedures and a problem-specific objective function, exhibiting excellent convergence efficiency and solution quality in quantum optimization problems of various sizes. Particularly, it may alleviate the barren plateau problem, a pervasive issue in variational quantum algorithms, and surpasses the performance of the VQE-SA method, an approach designed specifically to avoid barren plateaux. Additionally, the capability of VGON to identify degenerate ground states of quantum many-body models underscores its efficacy in addressing problems with multiple optima. Beyond the quantum world, generative models are emerging as powerful tools in the field of optimization problems. For instance, diffusion models are now being utilized for combinatorial optimizations~\cite{sanokowski2024diffusion}. Due to the flexible designs, we also envisage VGON and such algorithms to complement each other in addressing a broader spectrum of optimization challenges.
\\

\textbf{Acknowledgements}
This work is supported by the Sichuan Provincial Key R\&D Program (2024YFHZ0371), the National Natural Science Foundation of China (62250073, 62272259, 62332009) and the National Key R\&D Program of China (2021YFE0113100). The authors would like to thank Abolfazl Bayat, Dongling Deng, Chu Guo, Zhengfeng Ji, Damian Markham, Miguel Navascu\'{e}s and Ying Tang for helpful comments.
\\

\textbf{Code availability}
The complete code of this study is openly accessible via the GitHub repository \url{https://github.com/zhangjianjianzz/VGON}.


\clearpage

\renewcommand{\figurename}{Supplementary Fig.}
\setcounter{figure}{0}

\renewcommand{\tablename}{Supplementary Table}
\setcounter{table}{0}

\newcommand{\continuousappendix}{%
  \appendix
  \renewcommand{\theequation}{S\arabic{equation}} 
}

\section*{Appendix}
\subsection*{Reaching the quantum limit of a many-body contextuality witness}\label{appendix:tdvp}
Contextuality, a variant of quantum nonlocality when space-like separation can not be guaranteed, can be certified by the violation of a kind of inequalities called contextuality witnesses. For example, a typical contextuality witness is given by~\cite{Yang20221DContextuality}
\begin{equation}
    \begin{aligned}			
         & -4 \langle O_b^1 \rangle + 2 \langle O_a^1O_a^{2} \rangle + 2 \langle O_a^1O_b^{2} \rangle -2 \langle O_b^1O_a^{2} \rangle + 2 \langle O_b^1O_b^{2} \rangle \\ & + \langle O_a^1O_b^{3} \rangle - \langle O_b^1O_a^{3} \rangle \geq -4,
    \end{aligned}
    \label{322 cw}
\end{equation}
where $\{ \langle O_x^1 \rangle: x\in \{a,b\} \}$, $\{ \langle O_x^1O_y^2 \rangle: x,y\in \{a,b\} \}$ and $\{ \langle O_x^1O_y^3 \rangle: x,y\in \{a,b\} \}$ are the expectations of single-site correlator, nearest-neighbor and next-to-nearest neighbor two-point correlators, respectively. 

Ref.~\cite{Yang20221DContextuality} shows that for a given contextuality witness, the strongest violation that a quantum many-body system exhibits can be characterized as below. First, we transform the contextuality witness into a 1D infinite translation-invariant (TI) Hamiltonian with the fixed couplings being the same as the coefficients in the contextuality witness. Second, we choose the optimal local observables for the Hamiltonian such that the ground state energy density (GSED) is the lowest.

For example, we can parameterize the local observables $O_x:x\in\{a,b\}$ as
\begin{equation}
    O_x(\bm \theta_x) = (e^{\sum_{j=1}^{m}\theta_{xj}S_j})\Lambda_x(e^{\sum_{j=1}^{m}\theta_{xj}S_j})^{T},
\end{equation}
where $\Lambda_x$ is a diagonal matrix with entries being $\pm 1$, $\{S_j\}$ are the basis of the space of skew-symmetric matrices with the dimension of $m=(d^2-d)/2$, $d$ is the local dimension, and  ${\bm \theta_x} \equiv (\theta_{x1}, \theta_{x2},\dots,\theta_{xm})$ are real scalars. All the parameters combined are denoted as ${\bm{\theta}} \equiv (\bm \theta_a, \bm \theta_b)$. Then the Hamiltonian corresponding to contextuality witness~\eqref{322 cw} can be expressed as 
\begin{equation}
    \begin{aligned}			
         H(\bm{\theta})
         =\sum_{i=1}^{\infty} &-4 O_b^i(\bm \theta_b)+ 2O_a^i (\bm \theta_a) O_a^{i+1} (\bm \theta_a) + 2 O_a^i(\bm \theta_a)O_b^{i+1}(\bm \theta_b)\\ 
         &- 2O_b^i(\bm \theta_b)O_a^{i+1}(\bm \theta_a)  + 2 O_b^i(\bm \theta_b)O_b^{i+1}(\bm \theta_b) \\
         & + O_a^i(\bm \theta_a)O_b^{i+2} (\bm \theta_b) - O_b^i(\bm \theta_b)O_a^{i+2}(\bm \theta_a),
    \end{aligned}
    \label{322 hamiltonian}
\end{equation}
where $O_a^i(\bm \theta_b)$ and $O_b^i(\bm \theta_b)$ are two dichotomic observables on site $i$. We denote its GSED by $e(H({\bm{\theta}}))$, which can be calculated by the time-dependent variational principle (TDVP) algorithms \cite{PhysRevLett.107.070601, PhysRevB.88.155116}.

As a result, finding the strongest violation to the contextuality witness in Eq.~\eqref{322 cw} is now equivalent to solving the following minimization problem:
\begin{itemize}
    \item \textbf{Objective function}: $h(\bm{\theta}) = e(H(\bm \theta))$
    \item \textbf{Parameter space}: $\{\bm{\theta}\in\mathbb{R}^{d^2-d}\}$
\end{itemize}

In fact, by a modified version of the Navascués-Pironio-Acín (NPA) hierarchy~\cite{Yang20221DContextuality}, a lower bound for the lowest GSED of $H(\bm{\theta})$ has been obtained to be -4.4142134689. However, whether there is any infinite TI quantum many-body Hamiltonian can achieve this bound is still unknown. Using stochastic gradient descent (SGD), Ref.~\cite{Yang20221DContextuality} reports an infinite TI model that the corresponding GSED is $-4.4142131947$, which has a physical dimension $d = 5$ and a bond dimension $D = 5$. 

Combining these two results together, we can pin down the lowest GSED of $H(\bm{\theta})$ to the seventh significant digit.

We apply VGON to the above optimization problem. The model we choose contains a 2-layer encoder network with sizes [8, 4], a latent space with dimension 2, and a 3-layer decoder network with sizes [4, 8, 16]. In addition, we set the batch size as 2 and the learning rate as 0.005. It turns out that among all the outputs generated by VGON, 100\% can achieve a GSED of -4.4142134, improving the precision to eight significant digits.

\subsection*{Finding the optimal state for entanglement detection}\label{appendix:locc}

Suppose Alice and Bob are separated physically and want to determine whether a shared quantum state is entangled or not. For this, they play the prepare-and-measure entanglement detection game, where their goal is to design powerful measurement protocols such that the probability that they make mistakes is minimized. 

In a typical hypothesis test, all errors can be classified into two categories: type-I error (false-positive statistical error, i.e., concluding "Yes" when the state is not entangled) and type-II error (false-negative statistical error, i.e., concluding "No" when the state is entangled).

On a given quantum state $\rho$, in order to compare the power of local operations and one-way classical communication (1-LOCC) protocols and that of local operations (LO) ones on this problem, we can first fix the type-I error probability to be $p_1$, and then compare the minimal type-II error probability $p_2^*$ that these two kind of protocols can achieve. Furthermore, it has been known that $p_2^*$ can be calculated by the following semidefinite programming (SDP) optimization problems:  
\begin{equation}
    \begin{aligned}
        \min \quad  & p_2 \\
        \text{subject to }\quad & {\rm tr}(M_N\rho)=p_2,\\
        & p_1 \mathbb{I}-M_Y \in \mathcal{S}^*, \\
        & P \in \mathcal{P}^{\{LO,1-LOCC\}}.
    \end{aligned}
\end{equation}
Here $M_{Y(N)}$ is the positive operator-valued measure (POVM) operator with the outcome "Yes" (or "No") and can be expressed as a linear combination of variable $P$ and the measurement operators  $\{A_x^a\}$ ($\{B_y^b\})$ implemented in Alice's (Bob's) side, i.e.,
\begin{align*}
    M_{Y(N)}(P) = \sum_{x,y,a,b}P(x,y,\gamma=Y(N)|a,b)A^a_x\otimes B^b_y,
\end{align*}
where $x$ ($y$) denotes the label of the measurement settings, and $a$ ($b$) denotes the corresponding outcomes.
    
In addition, for different protocols the set of feasible optimization variables $P \in \mathcal{P}^{\{LO,1-LOCC\}}$ is restricted by different physical constraints. In LO protocols, the variable $P$ is required to satisfy
\begin{align*}
    \sum_{\gamma}P(x,y,\gamma|a,b)=P(x,y),\quad \sum_{x,y}P(x,y)=1,
\end{align*}
while for 1-LOCC protocols, we suppose that Alice makes the measurement first and then sends her measurement setting $x$ and outcome $a$ to Bob, making $P$ satisfy
\begin{align*}
    \sum_{\gamma}P(x,y,\gamma|a,b)=&P(x,y|a),\quad \sum_{y}P(x,y|a)=P(x),\\
    &\sum_xP(x)=1.
\end{align*}
  
Meanwhile, recall that the separable set is characterized by a hierarchical manner \cite{dohertyDistinguishingSeparableEntangled2002,PhysRevA.69.022308}. Taking the first level of hierarchical characterization into consideration, the constraint $p_1 \mathbb{I}-M_Y \in \mathcal{S}^*$ is dually equivalent to that the semidefinite positive matrices $M_0$ and $M_1$ satisfy
  \begin{align*}
        p_1 \mathbb{I}-M_Y = M_0+M_1^{T_B},
    \end{align*}
  where $T_B$ denotes the partial transpose with respect to Bob's subsystem.

  As a result, when quantifying the advantage of 1-LOCC protocols over LO ones in detecting the entanglement of $\rho$, we can focus on the following two SDP optimization problems and compute the gap between their solutions: 
\begin{equation}
    \begin{aligned}\label{eq:1-level_sdp}
        \min_{P, M_0, M_1}\quad  & p_2 \\
        \text{subject to }\quad & {\rm tr}(M_N\rho)=p_2 \\
        & p_1 \mathbb{I}-M_Y = M_0+M_1^{T_B},\ M_0,M_1\succcurlyeq 0 \\
        & M_Y = \sum_{x,y,a,b}P(x,y,\gamma=Y|a,b)A^a_x\otimes B^b_y \\
        & M_N = \sum_{x,y,a,b}P(x,y,\gamma=N|a,b)A^a_x\otimes B^b_y \\
        & P(x,y,\gamma|a,b)\ge 0,\ P \in \mathcal{P}^{\{LO,1-LOCC\}}
    \end{aligned}.
\end{equation}

Additionally, to make a fair comparison, in both 1-LOCC and LO protocols Alice and Bob choose the same set of quantum measurement settings as below:
    \begin{align*}
        &A_1^1=|0\rangle,\ A_1^2=|1\rangle,\ A_1^3=|2\rangle,\\
        &A_2^1 = \frac{1}{\sqrt{3}}(|0\rangle+e^{-i \frac{2 \pi}{3}}|1\rangle+e^{-i \frac{-2 \pi}{3}}|2 \rangle), \\
        &A_2^2 = \frac{1}{\sqrt{3}}(|0\rangle+e^{-i \frac{-2 \pi}{3}}|1\rangle+e^{-i \frac{2 \pi}{3}}|2 \rangle), \\
        &A_2^3 = \frac{1}{\sqrt{3}}(|0\rangle+|1\rangle+|2\rangle), \\
        &A^1_3 =\frac{1}{\sqrt{2}}(|1\rangle-|2\rangle),\ A^2_3 =|0\rangle,\ A^3_3 =\frac{1}{\sqrt{2}}(|1\rangle+|2\rangle).\\
    \end{align*}

In this work, we would also like to observe the advantage of 1-LOCC protocols over LO ones in quantum experiments. However, for a typical quantum state $\rho$ the above gap is very small, which makes the experimental observations very challenging, considering the imperfections of instruments and experimental noises. Therefore, we need to search for a quantum state that maximizes the above gap. Meanwhile, due to the limitations in experimental state preparations, we have to make the optimization among experiment-friendly states only. 

We employ both SGD and VGON to search for the optimal experiment-friendly pure state to exhibit the advantage of 1-LOCC protocols. Our results show that the VGON model is capable of generating the best pure states in approximately two hours, whereas it takes SGD over two months to achieve the same results. This sharp comparison highlights the significant advantage of VGON over the SGD method in tackling this problem.

Rigorously speaking, however, we cannot ensure that the optimal advantage is achieved by pure states, hence we also run VGON to search the maximal gap in the submanifold of mixed states space. Interestingly, numerical calculations show that this does not increase the observed gap further, implying that the largest gap is indeed achieved by pure quantum states. 
 
\subsubsection*{The pure state case}\label{appendix:pure}
    In a typical quantum laboratory, usually only a fraction of all quantum states can be prepared conveniently. Taking the photonic platform as an example, photons generated by the source interfere with each other and produce a quantum state expressed as $\sum_i c_i|i \rangle$, where $c_i$ are complex numbers satisfying $\sum_i |c_i|^2=1$. Then it can be transformed to $|\psi \rangle=\sum_i c_i|ii \rangle$ through spontaneous parametric down-conversion (SDPC). Specifically, when $|\psi\rangle$ is a qutrit-qutrit quantum system, it can be parameterized as~\cite{OptimalDec,Hu2018}
	\begin{align*}
		|\psi \rangle = \sin \frac{\theta}{2} \cos \frac{\phi}{4} e^{im} |00 \rangle +
		\sin \frac{\theta}{2} \sin \frac{\phi}{4} e^{in} |11 \rangle+
		\cos \frac{\theta}{2}|22 \rangle,
	\end{align*}
	where $\phi,m,n \in [0,2\pi)$, and $\theta \in [0,\pi]$.
    Then two local unitaries denoted by $U_A, U_B$ can be applied on the two subsystems, resulting in the quantum state
	\begin{align*}
	\rho_{\rm exp}=(U_A\otimes U_B)|\psi\rangle\langle\psi|(U_A^{\dagger}\otimes U_B^{\dagger}).
	\end{align*}
    Here $U_A$ ($U_B$) can be parameterized by a set of $3^2=9$ linearly-independent skew-Hermitian matrices $\{T_j\}$ \cite{hylandLearningUnitaryOperators2017}, i.e.,
	\begin{align*}
		U_A\ (U_B) = \exp\left(\sum_{j = 1}^{9}\lambda_j T_j\right),
	\end{align*}
	where $\lambda_j$'s are $9$ real numbers, denoted as $\bm{\lambda}_A$ ($\bm{\lambda}_B$).
 
 Then the problem is formulated as the following maximization problem:
	\begin{itemize}
		\item \textbf{Objective function}: $h(\rho_{\rm exp},p_1)={p_2^{LO}}^*(\rho_{\rm exp},p_1)-{p_2^{1-LOCC}}^*(\rho_{\rm exp},p_1)$
		\item \textbf{Parameter space}: $\{e_1,m,n,\phi,\theta \in \mathbb{R},\ \bm{\lambda}_A,\bm{\lambda}_B\in\mathbb{R}^{9}\}$.
	\end{itemize}
    Here we set $p_1=({\rm tanh}(e_1)+1)/2$, and ${p_2^{LO}}^*$ and $\ {p_2^{1-LOCC}}^*$ are computed by solving the two SDPs in Supplementary Eq.~(\ref{eq:1-level_sdp}) respectively.

    To fully test the performance of SGD on this problem, we run this algorithm for 79,663 different initial states, which are sampled from the parameter space according to the distribution $\mathcal{N}(\bm{0,I})$. The results are listed in Supplementary Figure~\ref{fig:UXU_all}(a), where the green and the blue dots present the gaps for the initial states and the optimized states, respectively. It turns out that GD gets stuck in local minima easily, and only 1.52\% of the initial states achieve a gap larger than 0.08. The largest gap observed is 0.0837. Lastly, we would like to stress that the above calculations take more than two months on a desktop-grade computer.
    \begin{figure}[!ht]
        \centering
        \includegraphics[width=0.98\columnwidth]{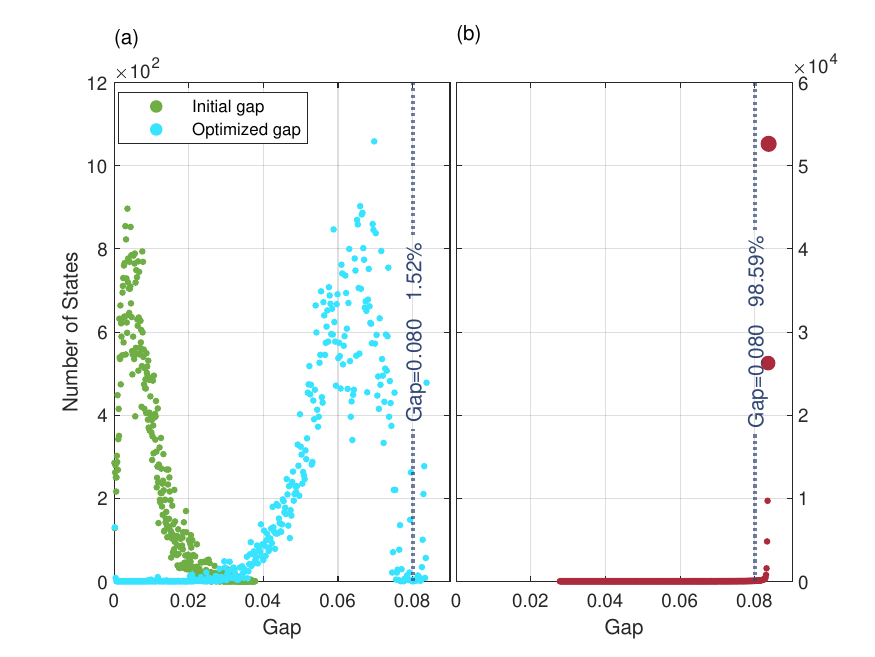}
        \caption{Distributions of gaps achieved by SGD and VGON. (a) Distributions of gaps obtained by SGD in two months. The green and blue dots present the gaps achieved by the 79,663 initial states of SGD and those by the corresponding optimized states, respectively. 1.52\% of them achieve a gap larger than 0.08, which is represented by the dark-blue dotted line. The obtained maximal gap is roughly 0.0837. (b) Distributions of the gaps optimized by VGON in two hours. Remarkably, 98.59\% of them exceed 0.08, and 52.657\% of them even fall within the range $[0.0836, 0.0837]$.}
        \label{fig:UXU_all}
    \end{figure}

    Subsequently, the same problem is addressed by VGON. The architecture of the VGON model consists of a 3-layer encoder network with sizes [512, 256, 128], a 3-layer decoder network with sizes [128, 256, 512], and a 2-dimensional latent space connecting the encoder and decoder components. The training is conducted with a batch size of 6, and an exponential decaying learning rate $lr_i$ at iteration $i$, where $lr_{i}=0.99 \times lr_{i-1}$, and $lr_{1}=0.001$. Once trained, the VGON model generates 100,000 quantum states as the output, and the gaps achieved by these quantum states are listed in Supplementary Figure~\ref{fig:UXU_all}(b), where we can see that the performance of VGON overwhelmingly surpasses that of the GD method.  Specifically, 98.59\% of the quantum states that VGON generates exhibit a gap larger than 0.08, and 52.657\% of them even fall within the range  $[0.0836, 0.0837]$, which is also the maximal gap found by SGD.

    \subsubsection*{The mixed state case}\label{appendix:mix}
    
    A subset of mixed quantum state $\rho\in\mathcal{H}^3\otimes\mathcal{H}^3$ can be parameterized by 
	\begin{align}
		\rho=U\Sigma U^{\dagger},
	\end{align}
    where $\Sigma$ is a $9\times 9$ diagonal matrix whose diagonal entries are nonnegative and sum to $1$, and $U$ is a unitary matrix that can be parameterized by a set of $9^2-1$ generalized Gell-Mann matrices $\{T_j\}$~\cite{Bertlmann2008}, i.e.,
    \begin{align*}
        U=\exp\left(i\sum_{j=1}^{9^2-1}\lambda_jT_j\right),
    \end{align*}
    where $\lambda_j$'s are $9^2-1$ real numbers, denoted as $\bm{\lambda}\in\mathbb{R}^{9^2-1}$. 
    
    We search for the mixed quantum state that achieves the largest gap with different methods including VGON. In order to facilitate the parameter update during the optimization process, we again set $p_1$ as $({\rm tanh}(e_1)+1)/2$, and write $\Sigma$ as ${\rm diag}(\sigma_1^2,\cdots,\sigma_9^2)$. If we let $\bm{\sigma}=(\sigma_1,\cdots,\sigma_9)$, then it holds that $\|\bm{\sigma}\|_2=1$. In this way, the problem is formed as  the following maximization problem
    \begin{itemize}
        \item \textbf{Objective function}: $h(\rho,p_1)={p_2^{LO}}^*(\rho,p_1)-{p_2^{1-LOCC}}^*(\rho,p_1)$
        \item \textbf{Parameter space}: $\{e_1\in \mathbb{R}, \bm{\lambda}\in\mathbb{R}^{9^2-1}, \bm{\sigma}\in\mathbb{R}^9:\|\bm{\sigma}\|_2=1\}$
    \end{itemize}

    For such a task, the chosen VGON model comprises a 4-layer encoder network and a 3-layer decoder network with sizes [1024, 512, 256, 128] and [128, 256, 512] respectively. We maintain the same latent space dimension and the same learning rate as those used in the training for pure states. In addition, we train the VGON model with a batch size of 3. Our results are depicted in Supplementary Figure~\ref{fig:mix}(a), which indicates that the VGON model is exceptionally adept at performing this task. Notably, 99.98\% of the parameter sets generated by the VGON model manifest a gap of 0.07, and 99.32\% of them even surpass a gap of 0.08. Particularly, as shown in Supplementary Figure~\ref{fig:mix}(b), when starting from a variational submanifold of the space of all mixed states, VGON always identified an almost pure state maximizing the gap, where the minimum achieved purity is 0.9993, and 96.49\% of the states have a purity greater than 0.9999. This shows the excellent capability of VGON in identifying qualified quantum states from the complex quantum state landscape without being stuck in local minima. 
    
    \begin{figure}[!ht]
        \centering
        \includegraphics[width=0.98\columnwidth]{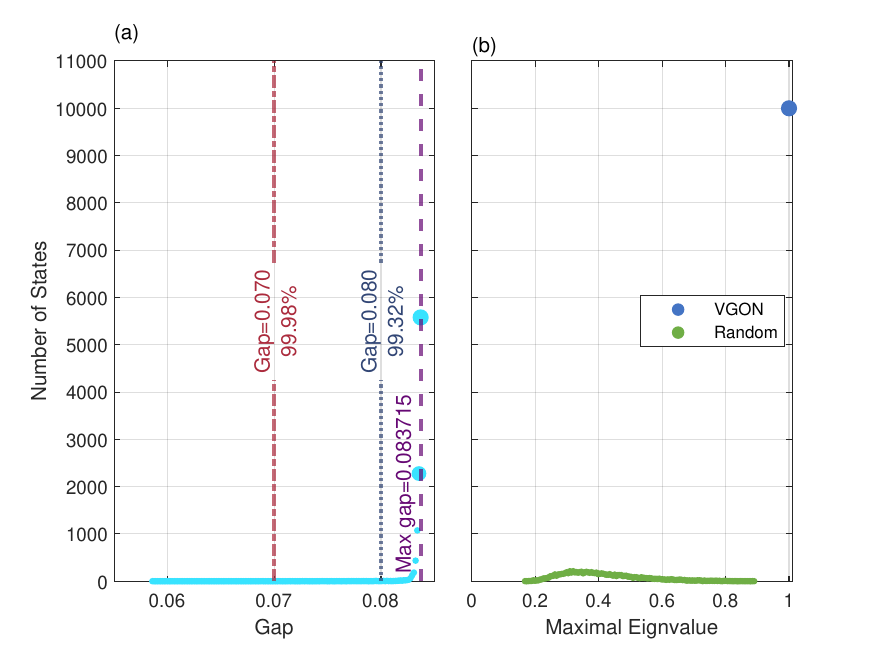}
        \caption{Numerical results for exhibiting the advantage of 1-LOCC over LO. (a) Distribution of the advantage brought by the VGON. (b) Distributions of the maximal eigenvalues for random training states and those generated by the VGON.}
        \label{fig:mix}
    \end{figure}
    
    \subsubsection*{Comparison between VGON and various global optimization algorithms}\label{appendix:comparison}

    For comparison purposes, we also apply seven other global optimization algorithms to this problem. For gradient-based algorithms, we choose GlobalSearch and Multistart \cite{globalsearch}, which both run repeatedly in parallel and attempt to find multiple local solutions with the help of certain strategies for choosing starting points. For gradient-free algorithms, we focus on Genetic Algorithm (GA) \cite{10.5555/522098}, Particle Swarm Optimization (PSO) \cite{7869491}, Simulated Annealing (SA) \cite{simulatedannealing}, Pattern Search (PS) \cite{patternsearch} and Surrogate optimization (SO) \cite{Gutmann:2001aa}. Since these algorithms are gradient-free, the updates of parameters are relatively easy to compute, while the optimization directions may not be accurate. 
    
    Additionally, Multilayer Perceptron (MLP) is also employed, which is a simple artificial neural network consisting of multiple fully connected layers.
    
    Using the default settings of the programs implemented by MATLAB, all the global optimization algorithms are executed on a computer with an Intel i9-12900KS Core and a RAM of 128 GB for 24 hours. Meanwhile, an MLP model with a batch size of 10, consisting of a 7-layer network with each layer containing 90 neurons is also trained 10 times with the other configuration remaining the same as that of VGON. The numerical results are shown in Supplementary Table~\ref{Tab:compare}, where we can clearly see that the performance of VGON exceeds that of all the other methods.

    \begin{table}[ht]
    \centering
    \caption{Comparisons of VGONs with seven global optimization algorithms and MLP.}
    \setlength{\tabcolsep}{1mm}{
    \begin{tabular}[t]{lcc}
        \hline
        \hline
        Algorithm&Maximal Optimized Gap&Run time\\
        \hline
        GlobalSearch&1.2e-07&24 hours\\
        Multistart&0.0726&24 hours\\
        GA&0.0779&24 hours\\
        PSO&0.0348&24 hours\\
        SA&0.0067&24 hours\\
        PS&0.0717&24 hours\\
        SO&0.0412&24 hours\\
        MLP&0.0384 (mean)&3,000 iterations\\
        \hline
        VGON&0.0837&3,000 iterations\\
        \hline
        \hline
    \end{tabular}}\label{Tab:compare}
    \end{table}

\subsection*{Alleviating the effect of barren plateaux in variational quantum algorithms}\label{appendix:barren_plateaus}

The aim of the variational quantum eigensolver (VQE) is to approximate the ground state $|\psi_{G}\rangle$ of a target Hamiltonian $H$ with a variational wave function

\begin{align*}
  |\psi(\bm{\theta})\rangle=U(\bm{\theta})|00\cdots0\rangle,
\end{align*}
where variational parameters $\bm{\theta}\in \mathbb{R}^{M}$, and $M$ is determined by the practical ansatz of the parameterized quantum circuit (PQC). To ensure a close approximation to the ground state,  $\bm{\theta}$ is iteratively updated through a classical computer by a gradient descent algorithm aiming at minimizing the energy, which forces $|\psi(\bm{\theta})\rangle$ to be close to the ground state $|\psi_{G}\rangle$. More concretely, the optimization problem is formed as the following minimization problem:
\begin{itemize}
    \item \textbf{Objective function}: $h(\bm{\theta})=\langle\psi(\bm{\theta})|H|\psi(\bm{\theta})\rangle$
    \item \textbf{Parameter space}: $\{\bm{\theta}\in\mathbb{R}^{M}\}$
\end{itemize}
However, gradient-based optimization methods often encounter a notable challenge called barren plateaus (BPs), which are characterized by exponentially vanishing gradients. This issue typically emerges from random initializations of parameterized unitaries that admit the statistics of a unitary 2-design \cite{Harrow:2009aa}. 

In this section, we first apply PQCs on large-scale quantum problems to replicate the BP phenomenon, where the magnitude of parameters, $M$, reaches up to $10^4$. Then we show that the VGON model can address this challenge very well, which not only showcases its capacity to handle large-scale optimization problems, but also highlights its critical advantage in overcoming the issue of gradient vanishing.

\subsubsection*{The $Z_1Z_2$ model}\label{appedix:bp_grads}
As a toy example, we first set the target Hamiltonian to be
\begin{align*}
    H = Z_1Z_2,
\end{align*}
i.e., a Pauli ZZ operator acting on the first and second qubits, and the corresponding ground energy is -1. This Hamiltonian was studied in Ref. \cite{McCleanBarrenPlateau} to exhibit the existence of BPs. 

To approximate the ground states of $Z_1Z_2$, a hardware-efficient ansatz \cite{Kandala:2017aa}
\begin{align*}	U(\bm{\theta})=\prod_{l=1}^{L}U_{ENT}\left(\prod_{i=1}^NR_l^i(\theta_l^i)\right),
\end{align*}
is adopted. Here $\theta_l^i\in[0,2\pi)$ are variational angles, and all the $L\times N$ such angles combined is denoted as $\bm{\theta}$. The rotations $R_l^i(\theta_l^i)=\exp(-\frac{\rm{i}}{2}\theta_l^iG_l^i)$ have random directions given by $G_l^i\in\{\sigma_x,\sigma_y,\sigma_z\}$. $U_{ENT}$ is an entangling unitary operation consisting of two-qubit nearest-neighbor controlled-Z (CZ) gates with periodic boundary conditions. $L$ and $N$ correspond to the numbers of the layers and qubits of $U(\bm{\theta})$ respectively. The structure of $U(\bm{\theta})$ for the $Z_1Z_2$ model is schematically shown in Supplementary Figure~\ref{fig:circ_structure}.
\begin{figure}[!ht]
\centering
\includegraphics[width=0.98\columnwidth]{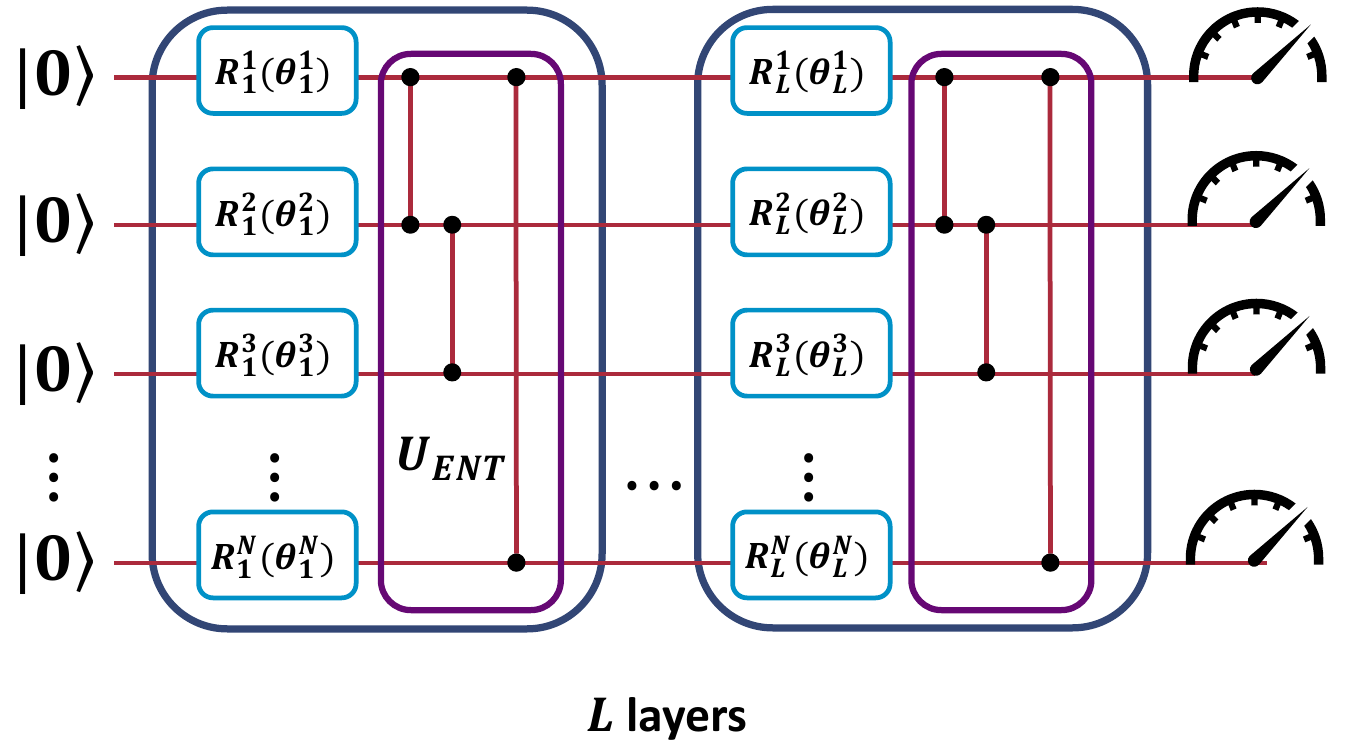}
\caption{Structure of the parameterized quantum circuit for the $Z_1Z_2$ model. Each dark-blue wireframe represents a layer of the circuit consisting of single-qubit rotations represented by light-blue boxes and an entangling unitary operation $U_{ENT}$ represented as a purple box, where entangling CZ gates are shown by lines. The measurements at the end are used to estimate the energy of the trial state.}
\label{fig:circ_structure}
\end{figure}

Let the numbers of qubits and layers be $20$ and $400$ respectively. To approximate the ground states of the $Z_1Z_2$ model by VQE, the variable $\bm{\theta}\in\mathbb{R}^{8,000}$ is uniformly initialized from the parameter space, i.e., $\theta_l^i\in[0,2\pi)$, and then updated iteratively by the Adam optimizer \cite{kingma2017adam} to minimize $h(\bm{\theta})$. After 300 iterations, the obtained energy is -0.0068, which is actually far from the ground energy -1. Particularly, the variance of $\{\partial_{{\theta_l^i}}h(\bm{\theta})\}$ decreases from $4.4999\times 10^{-7}$ right after the initialization to $1.3351\times 10^{-9}$ at the last iteration, which indicates that VQE suffers from BPs heavily, hence can hardly achieve the ground energy. The dark-blue boxes in Supplementary Figure~\ref{fig:boxplot_grads_z1z2} and the dark-blue line in Supplementary Figure~\ref{fig:history_z1z2} plot these numerical results, which also match the results reported in Ref. \cite{McCleanBarrenPlateau}. 

Meanwhile, several promising strategies for avoiding BPs have been proposed and investigated, and small-angle initialization, denoted as VQE-SA, is a widely used technique \cite{PRXQuantum.3.010313, PRXQuantum.2.040309, PRXQuantum.3.020365}. The VQE-SA method tries to initialize $\bm{\theta}$ near the zero vector, thus differing the statistics of $U(\bm{\theta})$ from a 2-design to avoid BPs. 

We now apply VGON to solve the same problem. Since VGON is designed to map a bunch of different initial values of the variable  $\bm{\theta}$ to the optimal ones, it may break improper initializations of random quantum circuits that lead to BPs. Besides, VGON contains a sampling procedure in the latent space to bring randomness, which may also help to maintain larger gradients. Interestingly, we will show that this is indeed the case, and VGON not only can solve the $Z_1Z_2$ model very well, but also enjoys a remarkable advantage over the VQE-SA method in alleviating BPs.

To employ the VQE-SA method on this problem, the variable $\bm{\theta}$ is uniformly sampled from $\{\theta_l^i\in[0,0.01)\}$ as the starting point, and then updated iteratively to minimize $h(\bm{\theta})$. As for VGON, 1,200 $\bm{\theta}$'s are uniformly initialized from $\{\theta_l^i\in[0,2\pi)\}$ as inputs of the VGON model, whose structure contains a 4-layer encoder network with layer shape [256, 128, 64, 32], a latent space with dimension 3, and a 4-layer decoder network with layer shape [32, 64, 128, 256]. Set batch size to be 4, and the coefficient of the KL divergence to be $1/8$. With all the other configurations kept the same as the VQE method introduced above, we run both the VQE-SA method and VGON for 300 iterations. Note that the update for one batch in VGON counts for one iteration.

To {fairly} compare the gradient distributions of the two optimization algorithms,  we focus on $\{\partial_{{\theta_l^i}}h(\bm{\theta})\}$ for the VQE-SA method, and $\{\partial_{b_l^i}C(\bm{\phi,\omega})\}$ for VGON, where $b_l^i$ is the bias in the last layer of the decoder that contributes to the parameter $\theta_l^i$. {To explain why this is the case, notice that
\begin{align*}
    \theta_l^i=W^{(l,i)}\bm{x}+b_l^i,
\end{align*}
where $W^{(l,i)}$ represents a row of the weight matrix of the decoder's last layer and $\bm{x}$ is the output of the previous layer. Therefore, $\partial_{b_l^i}C(\bm{\phi,\omega})$ is influenced by the PQC $U(\bm{\theta})$ only, making the comparison with  $\partial_{{\theta_l^i}}h(\bm{\theta})$ quite fair.}

It turns out that right after the initialization, the variances of these two sets of gradients are $1.0695\times10^{-2}$ and $1.2742\times10^{-2}$, respectively, which are both five orders of magnitude larger than those in the original VQE method. When the optimizations are terminated, these variances eventually decrease to $7.8099\times10^{-14}$ and $6.0972\times10^{-6}$, respectively, with only VGON exhibiting a much larger variance magnitude than VQE. Supplementary Figure~\ref{fig:boxplot_grads_z1z2} illustrates the distribution of $\{|\partial_{{\theta_l^i}}h(\bm{\theta})|\}$ for the VQE-SA method with red boxplots, and that of $\{|\partial_{b_l^i}C(\bm{\phi,\omega})|\}$ for VGON with purple boxplots. As we can see, the absolute values of the gradients for the VQE-SA method and VGON are distributed more widely than those in VQE. Furthermore, a considerable part of these absolute values of gradients, especially at the initial stages, is several orders of magnitude larger compared to those in VQE, which is crucial for effectively decreasing the energies. 

\begin{figure}[!ht]
    \centering
    \includegraphics[width=0.98\columnwidth]{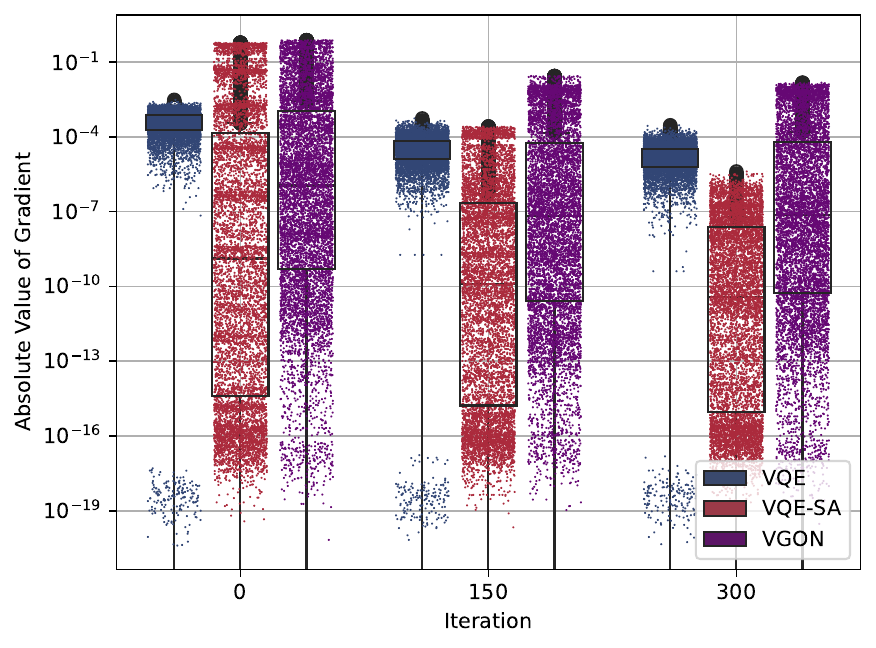}
    \caption{The absolute values of the gradients for the $Z_1Z_2$ model. The boxplots illustrate the distribution of $\{|\partial_{{\theta_l^i}}h(\bm{\theta})|\}$ for VQE (dark-blue) and the VQE-SA method (red), and the distribution of $\{|\partial_{b_l^i}C(\bm{\phi,\omega})|\}$ for the VGON (purple) at different iterations. Each boxplot displays the distribution based on a five-number summary: the minimum, the first quartile, the sampled median, the third quartile and the maximum. All other observed data points outside the minimum and maximum are plotted as outliers with black diamonds.}
    \label{fig:boxplot_grads_z1z2}
\end{figure}

In addition, Supplementary Figure~\ref{fig:history_z1z2} illustrates the energies at different iterations for the two methods. As depicted by the red dashed line, the VQE-SA method converges to -1 fast, which is exactly the ground energy. In VGON, the average (minimal) energy is represented by the green (purple) dashed line, which also decreases rapidly, and eventually achieves a minimum value of -0.9998 (-0.9999). Compared with the VQE-SA method, at the beginning the fluctuations in VGON are smaller, which means VGON converges to the ground energy faster in this stage. However, since VGON tries to map the uniformly random parameters to those centered around the optimal parameters, there remain weak fluctuations in the later stage, but it still manages to find the ground energy very well.

\begin{figure}[!ht]
    \centering
    \includegraphics[width=0.98\columnwidth]{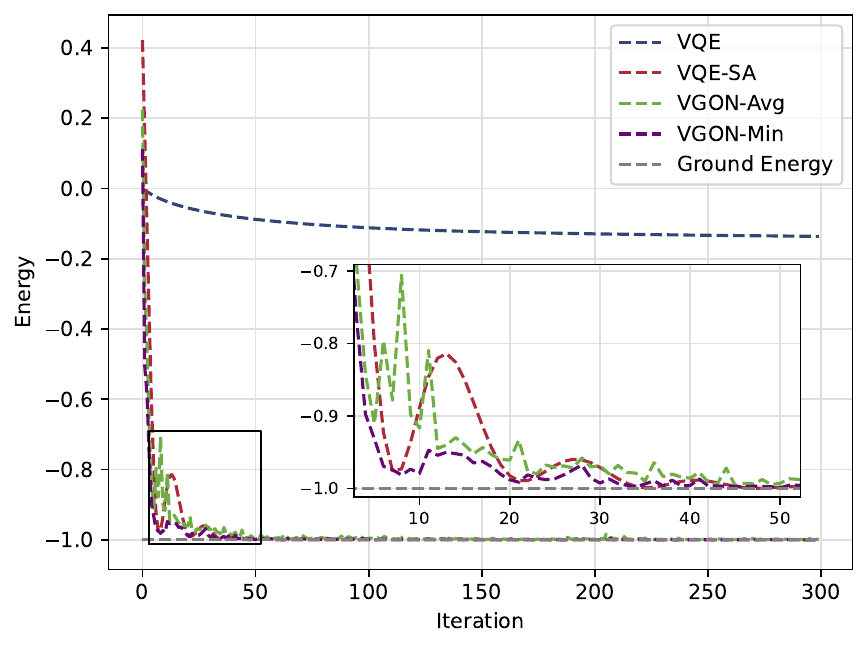}
    \caption{Energies of the $Z_1Z_2$ model at different iterations for different methods. The VQE (dark-blue) suffers from BPs and can hardly be optimized. The VQE-SA method (red) and VGON, whose average (green) and minimal (purple) energies in each batch are presented, converge to the ground energy quickly. The exact ground energy (gray) is -1.}
    \label{fig:history_z1z2}
\end{figure}

\subsubsection*{The Heisenberg XXZ model}\label{appendix:bp_xxz}

On the $Z_1Z_2$ model, VGON exhibits its advantage over VQE, but the separation between it and VQE-SA in terms of convergence speed is less obvious. To further investigate the advantage, we now move on to the Heisenberg XXZ model, and compare the performance of different methods from the perspective of the fidelity between the optimized state and the exact ground state.

The Hamiltonian of the Heisenberg XXZ model with periodic boundary conditions is given by
\begin{align*}
    H_{XXZ}=-\sum_{i=1}^{N}\left(\sigma_x^{i}\sigma_{x}^{i+1}+\sigma_y^{i}\sigma_{y}^{i+1}-\sigma_z^{i}\sigma_{z}^{i+1}\right),
\end{align*}
where $\sigma_{x,y,z}^i$ denote the Pauli operators at site $i$. For the number of qubits $N=18$, the exact average ground energy is -1.7828.

To find out the ground state of $H_{XXZ}$, a relatively more complex ansatz~\cite{ran2020-encoding}
\begin{equation}
    \begin{aligned}
        U'(\bm{\theta})=\prod_{l=1}^{L}\prod_{i=1}^{N-1}U_l^i(\bm{\theta}_l^i)
    \end{aligned}\label{eq:ansatz_2u}
\end{equation}
depicted in Supplementary Figure~\ref{fig:circ_structure_h232} (a) is applied. Here $L$ and $N$ are the numbers of the layers and qubits involved in $U'(\bm{\theta})$ respectively, and each $U_l^i(\bm{\theta}_l^i)$ is a univeral 2-qubit gate at the $l$-th layer acting on qubits $i$ and $i+1$, which is determined by $\bm{\theta}_l^i$ containing 15 rotation angles, as illustated in Supplementary Figure~\ref{fig:circ_structure_h232} (b).
\begin{figure}[t]
  \centering
  \begin{overpic}[width = 0.85\linewidth]{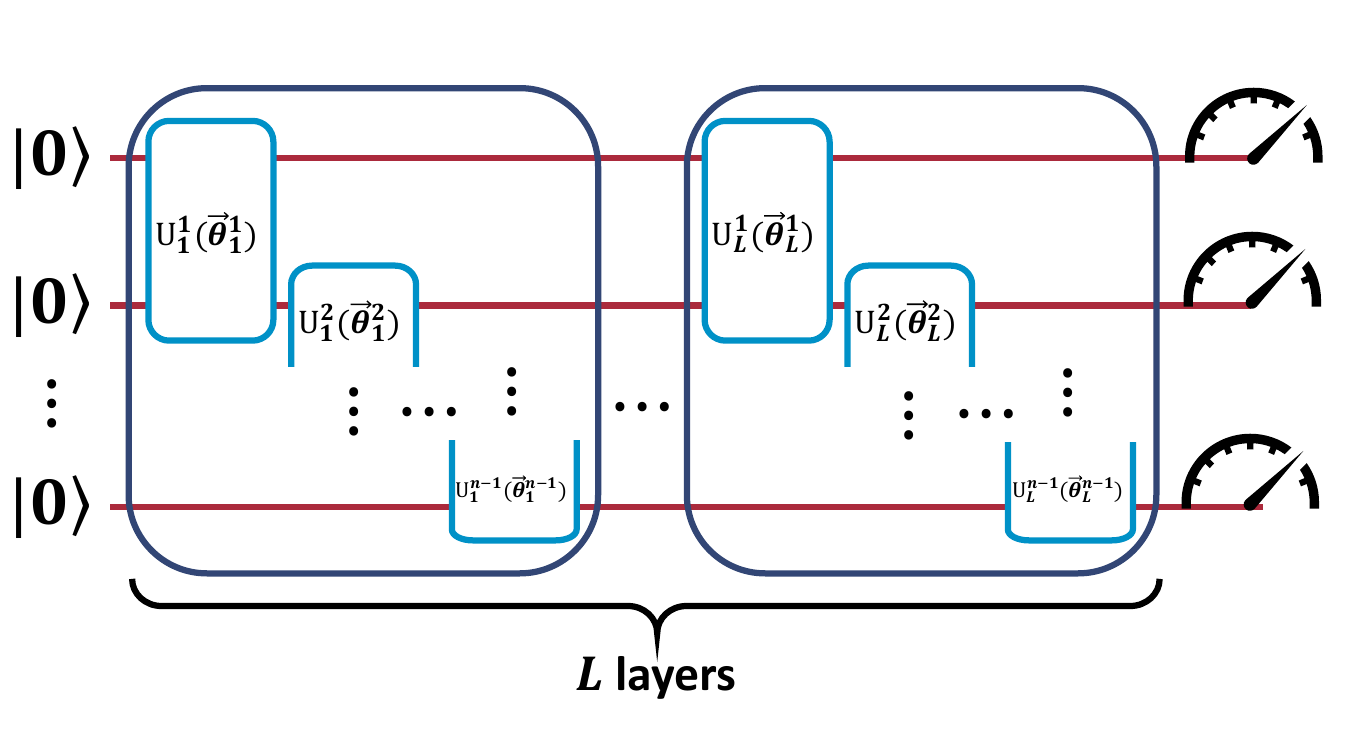}
      \put(-5, 47){\textbf{(a)}}
  \end{overpic}
  \begin{overpic}[scale=0.4]{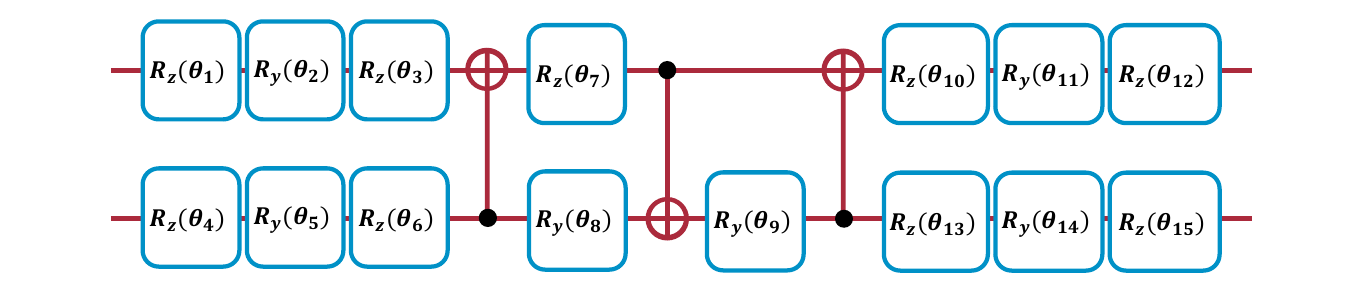}
      \put(-2, 23){\textbf{(b)}}
  \end{overpic}

  \caption{Structure of the parameterized quantum circuit for the Heisenberg XXZ model. (a) Each dark-blue box represents a layer of the circuit consisting of $n-1$ universal 2-qubit gate blocks. (b) Each universal 2-qubit gate block is decomposed into 15 rotation gates and 3 CNOT gates.}
  \label{fig:circ_structure_h232}
\end{figure}

Let $N$ and $L$ be 18 and 48, respectively. For VQE, $\theta_l^i$'s are firstly sampled from 0 to $2\pi$ uniformly as the starting point, and $\bm{\theta}\in\mathbb{R}^{12,240}$ is then updated iteratively to minimize $h(\bm{\theta})$. After repeating this optimization process 10 times, the mean value of the average energy after 1,000 iterations is found to be only -0.1367. Since the exact average ground energy is -1.7828, such a poor result indicates the strong impact of BPs.

When it turns to the VQE-SA method, a $\bm{\theta}$ is initialized uniformly from $\{\theta_l^i\in[0,0.01)\}$ as the starting point, and then it is updated iteratively to search for the ground state. As for VGON, 8,000 $\bm{\theta}$'s are uniformly initialized from $\{\theta_l^i\in[0,2\pi)\}$ as the inputs of the model, which contains a 7-layer encoder network with sizes [8192, 4096, 2048, 1024, 512, 256, 128], a latent space with dimension 100, and a 7-layer decoder network with sizes [128, 256, 512, 1024, 2048, 4096, 8192]. Set the batch size to 8, and the coefficient of the KL divergence to $0.1$. We run both of the two methods for 1,000 iterations with all the other configurations kept the same as VQE. To make a fair comparison, we repeat the whole process 10 times. Figure~3 in the main text illustrates the corresponding mean values and the 95\% confidence intervals of the energy densities and the fidelities between the optimized state and the ground state at different iterations, where we can clearly see the faster and more stable convergence of VGON than the VQE-SA method.

\subsection*{Identifying degenerate ground state space of quantum models}\label{appendix:degeneracy}

As evidenced earlier, the VGON model exhibits excellent capabilities in solving optimization problems with a single optimal solution. In this section, by solving a degenerate ground space we demonstrate that VGON also has the capability to effectively handle optimization problems with multiple optimal solutions.

To identify the degenerate ground space of a Hamiltonian $H$ with VGON, the objective function needs two pivotal components to steer the optimized quantum state $|\psi(\bm{\theta})\rangle$ towards diverse ground states. The first component utilizes a PQC $U(\bm{\theta})$ to generate the state $|\psi(\bm{\theta})\rangle=U(\bm{\theta})|00\cdots 0\rangle$, targeting the ground space. The second component integrates a cosine similarity measure into the optimization objective, aiming to enhance the diversity among the generated quantum states.

Specifically, for a batch of $S_b$ states $\{|\psi(\bm{\theta}_i)\rangle\}$, the mean energy is calculated by
\begin{align*}
    \bar{E}(\bm{\Theta})=\frac{1}{S_b}\sum_{i=1}^{S_b}\langle\psi(\bm{\theta}_i)|H|\psi(\bm{\theta}_i)\rangle,
\end{align*}
where $\bm{\Theta}=(\bm{\theta}_1, \bm{\theta}_2, \cdots, \bm{\theta}_{S_b})$. In addition, a penalty term for the objective function based on the cosine similarity is defined as
\begin{align*}
    \bar{S}_{\mathcal{C}_{S_b}^2}(\bm{\Theta})=\frac{1}{|\mathcal{C}_{S_b}^2|}\sum_{(i,j)\in \mathcal{C}_{S_b}^2}\frac{\bm{\theta}_i\cdot\bm{\theta}_j}{\Vert\bm{\theta}_i\Vert\Vert\bm{\theta}_j\Vert},
\end{align*}
where $\mathcal{C}_{S_b}^2$ represents the set of all 2-combinations pairs derived from the elements in $\{1,2,\cdots,S_b\}$, and $\Vert\cdot\Vert$ denotes the Euclidean norm. Eventually, the optimization objective is set as minimizing  of a combination of $\bar{E}(\bm{\Theta})$ and $\bar{S}_{\mathcal{C}_{S_b}^2}(\bm{\Theta})$ according to a trade-off coefficient $\gamma$, i.e.,
\begin{itemize}
\item \textbf{Objective function}: $h(\bm{\Theta}) = \bar{E}(\bm{\Theta})+\gamma \cdot \bar{S}_{\mathcal{C}_{S_b}^2}(\bm{\Theta})$
\item \textbf{Parameter space}: $\{\bm{\Theta}\in\mathbb{R}^{S_bM}\}$
\end{itemize}
In this section, we consider the ansatz expressed by Eq.~\eqref{eq:ansatz_2u}, hence $M$ equals $15(N-1)L$, where $N$ and $L$ are the number of qubits and layers in the circuit, respectively.

\subsubsection*{The Majumdar-Ghosh model}\label{appendix:degeneracy_MG}
The Majumdar-Ghosh (MG) model, a one-dimensional chain of interacting spins with next-nearest-neighbor interactions, is a classic example exhibiting substantial degeneracy under open boundary conditions, whose Hamiltonian is written as  
\begin{align*}
    H_{MG} = \sum_{i=1}^{N} \bm{\sigma}^i\cdot \bm{\sigma}^{i+1} + \bm{\sigma}^{i+1} \cdot\bm{\sigma}^{i+2}+ \bm{\sigma}^i \cdot\bm{\sigma}^{i+2},
\end{align*}
where $\bm{\sigma}^i = (\sigma^i_x, \sigma^i_y, \sigma^i_z)$ are Pauli operators at site $i$. In the MG model, the local 3-site term is a sum of 2-local swap operations, resulting in a 4-dimensional ground antisymmetric space. As the particle number grows, the ground space is determined by intersecting the added state space with the previous ground space, with dimensions of 4 for odd sizes and 5 for even sizes.

For the case that $N=10$, whose exact ground energy is -24, we set $L=4$, resulting in 36 universal 2-qubit gate blocks, and the batch size $S_b=50$. To balance the diversification of generated states with their eventual convergence to the ground space, the trade-off coefficient $\gamma$ is dynamically adjusted across different iterations using a step function that gradually decreases from 40 to 1.

The VGON model employed to tackle this task has a 4-layer encoder network with sizes [512, 256, 128, 64], a latent space with dimension [50], and a 4-layer decoder network with sizes [64, 128, 256, 512]. During the training procedure, a dataset randomly sampled from a uniform distribution on the interval $[0,1]$ is utilized. The hyperparameter $\beta$ serving as the coefficient of the KL divergence and the learning rate are set as 1 and 0.0014, respectively. The training is terminated upon reaching an energy value of -23.90. 

\begin{figure}[!ht]
    \centering
    \includegraphics[width=0.98\columnwidth]{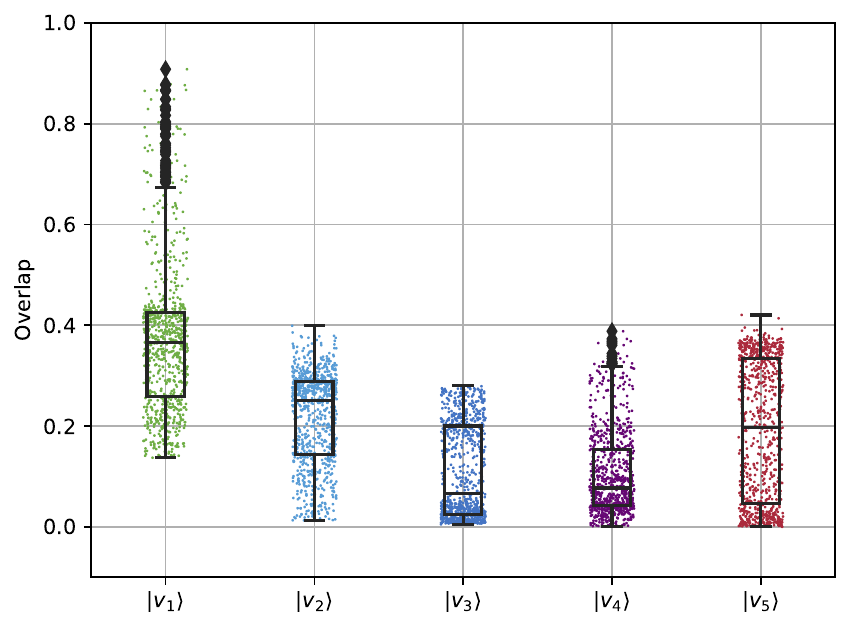}
    \caption{The overlaps between the generated states of VGON and the ground states. Each boxplot shows the degree of dispersion and skewness for the overlap with one ground state. Supplementary Table \ref{tbl:mg} lists the minimum, lower quartile, median, upper quartile, and maximum for each ground state.}
    \label{fig:degenerated_box}
\end{figure}

\begin{table}[!ht]
  \centering
  \caption{Details on the overlap boxes of VGON's output on the basis of the degenerate space of $H_{MG}$.}
  \fontsize{8}{10}\selectfont
  \setlength{\tabcolsep}{3pt}
  \begin{tabular}{c c c c c c}
  \hline
  \hline
  Basis &  Minimum & Lower Quartile & Median & Upper Quartile & Maximum\\
  \hline
  $|v_{1}\rangle$ & 0.1373 & 0.2581 & 0.3660 & 0.4252 & 0.9081\\
  $|v_{2}\rangle$ & 0.0129 & 0.1442 & 0.2509 & 0.2887 & 0.3992\\
  $|v_{3}\rangle$ & 0.0050 & 0.0254 & 0.0663 & 0.2002 & 0.2800\\
  $|v_{4}\rangle$ & 0.0007 & 0.0418 & 0.0772 & 0.1529 & 0.3882\\
  $|v_{5}\rangle$ & 0.0001 & 0.0463 & 0.1969 & 0.3335 & 0.4206\\
  \hline
  \hline
\end{tabular}
\label{tbl:mg}
\end{table}

\begin{figure}[!ht]
    \centering
    \includegraphics[width=0.98\columnwidth]{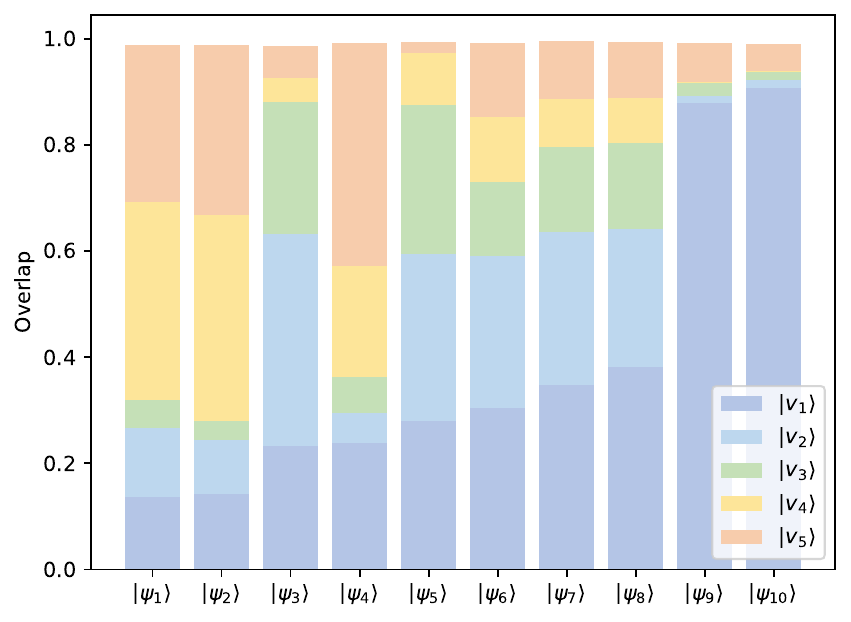}
    \caption{The overlaps between selected generated states and the degenerate space. Five bases of the degenerate space, represented by $|v_{1}\rangle-|v_{5}\rangle$, are determined through the exact diagonalization method. For each generated state $|\psi_i\rangle$, the corresponding bars with different colors represent the overlaps into different bases.}
    \label{fig:degenerated_bar_10}
\end{figure}

Among the 1000 generated states, 81.9\% achieve the energy threshold of -23.90. To examine the diversity of the generated states, we analyze the overlaps between the states achieving the above threshold and all the ground states, as shown in Supplementary Figure~\ref{fig:degenerated_box}. It can be clearly seen that the generated states exhibit significant diversity. To ensure that VGON's outputs effectively involve all the dimensions of the ground space, we set the overlap threshold to 0.001 and then analyze all the generated states. The results indicate that 81.4\% of the generated states meet both the energy and the overlap thresholds. Supplementary Figure~\ref{fig:degenerated_bar_10} exhibits ten such generated states, which not only illustrates the remarkable diversity of the solutions provided by VGON, but also demonstrates VGON's capability of identifying degenerate ground state spaces for quantum models effectively.

\subsubsection*{The 232 model}\label{appendix:degeneracy_232}

In this context, a model that exhibits a degeneracy of 2 for odd sites with open boundary conditions is considered, whose Hamiltonian is \cite{Yang20221DContextuality}
\begin{align*}
    H_{232}=\sum_{i=1}^{N}(2\sigma_{x}^i\sigma_{x}^{i+1}+\sigma_{x}^i\sigma_{y}^{i+1}-\sigma_{y}^i\sigma_{x}^{i+1}),
\end{align*}
where $\sigma_{x,y,z}^i$ represent the Pauli matrices at site $i$. Let the number of qubits $N$ be 11, the corresponding ground energy is $-20.7106$.

Consider $L=6$, resulting in the number of universal 2-qubit gate blocks being 60, and the batch size $S_b=50$. 
The configurations of the VGON model and the training procedure remain consistent with that in Appendix~\ref{appendix:degeneracy_MG}, except for setting the learning rate to 0.0015 and terminating the training upon reaching an energy value of -20.6106.

After training, 1,000 quantum states are generated by the VGON model, with 78.7\% demonstrating energies below $-20.6106$. As illustrated in Supplementary Figure~\ref{fig:rain_232}(a), the overlap distribution on each basis state, denoted as $\{|u_{1}\rangle,|u_{2}\rangle\}$, showcases a bimodal pattern, precisely reflecting the degree of degeneracy. Moreover, the analysis of fidelities between pairs of generated states, depicted in Supplementary Figure~\ref{fig:rain_232}(b), reveals values clustering around either 0 or 1. This indicates that the states generated by the VGON model are either identical or orthogonal to each other. Consequently, this affirms VGONs' capability to directly generate a complete set of orthogonal bases for this task.

\begin{figure}[!ht]
    \centering
    \includegraphics[width=0.98\columnwidth]{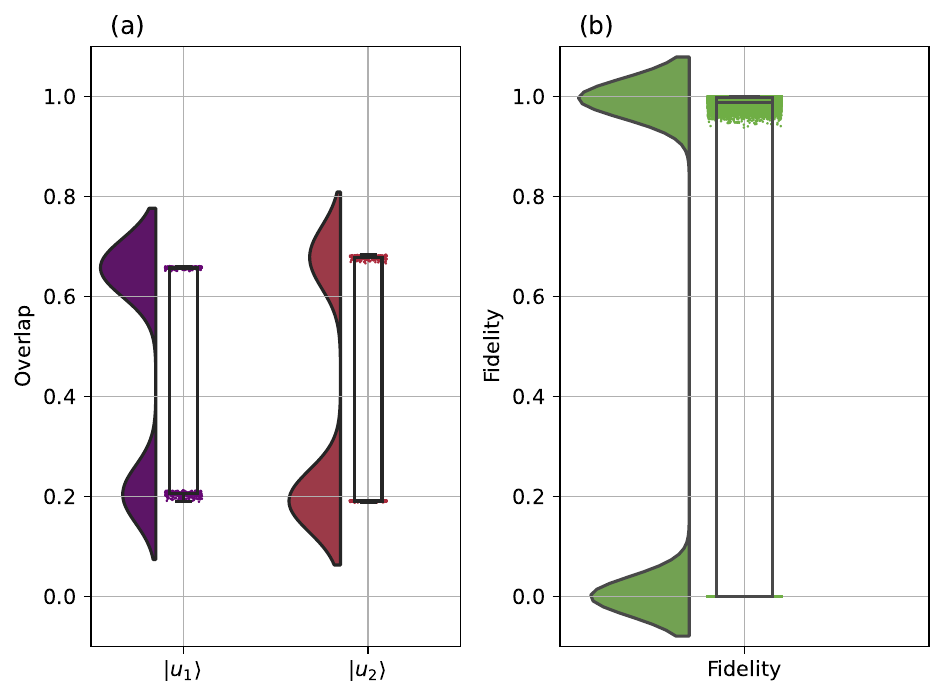}
    \caption{Distributions of the generated states for the 232-type model. (a) The overlap distributions for the basis states $|u_{1}\rangle$ and  $|u_{2}\rangle$, whose population densities plots are both bimodal, consistent with the degree of degeneracy. The minimum, lower quartile, median, upper quartile, and maximum for each basis state are presented in the first part of Supplementary Table \ref{tbl:232}.
    (b) The distribution of fidelities between pairs of generated states with energies below -20.6106. The population densities plot reveals a bimodal distribution, with pronounced peaks near 0 and 1. This pattern suggests that the states are predominantly either identical (fidelity close to 1) or orthogonal (fidelity close to 0) to each other. The statistical summary of the boxplot is shown in the second part of Supplementary Table \ref{tbl:232}.}
    \label{fig:rain_232}
\end{figure}

\begin{table}[!ht]
  \centering
  \caption{Boxplot details on overlap distributions for the basis of the degenerate space of $H_{232}$, and the distributions of fidelities between pairs of generated states.}
  \fontsize{8}{10}\selectfont
  \setlength{\tabcolsep}{1.2pt}
  \begin{tabular}{c c c c c c}
  \hline
  \hline
   &  Minimum & Lower Quartile & Median & Upper Quartile & Maximum\\
  \hline
  $|u_{1}\rangle$ & 0.1895 & 0.2063 & 0.6564 & 0.65820 & 0.6608\\
  $|u_{2}\rangle$ & 0.1873 & 0.1907 & 0.1913 & 0.6777 & 0.6845\\
  \hline
  Fidelity & $2.7511\times 10^{-15}$ & $4.8752\times 10^{-9}$ & 0.9882 & 0.9978 & 0.9998\\
  \hline
  \hline
\end{tabular}
\label{tbl:232}
\end{table}

\subsection*{Neural network settings for different tasks}
For clarity and brevity, we summarize the neural network hyperparameters used across these tasks in Supplementary Table~\ref{Tab:NNsettings}. It can be found that the choice of latent space dimension and KL coefficient should align with the problem scale and optimization landscape complexity. Moderate settings suffice for simpler tasks—such as LTI models or pure/mixed-state cases—and for models with smoother landscapes like the Z1Z2 model. In contrast, more complex systems, including the XXZ model with larger state spaces and the MG model prone to trapping in degenerate ground-state subspaces, demand higher latent dimensions and KL coefficient annealing.

Small latent dimensions suffice when the intrinsic dimensionality of optimal solutions is low, even for challenging problems like mixed states. However, as the solution space grows more complex or higher-dimensional, the latent space must expand accordingly.

\begin{table*}[htbp!]
  \centering
  \caption{Neural network settings for different tasks including encoder structure $E$, decoder $D$ batch size $S_b$, latent dimension $z$ and KL coefficients $\beta$.}
  \begin{tabular}{lccccc}
    \toprule
    \textbf{Tasks} & \textbf{$E$} & \textbf{$D$} & \textbf{$S_b$} & \textbf{$z$} & \textbf{$\beta$} \\
    \midrule
    LTI model      & [8,4]             & [4,8,16]        & 2                 & 2   & 1       \\
    Pure state case& [512,256,128]     & [128,256,512]   & 6                 & 2    & 1        \\
    Mixed state case& [1024,512,256,128]& [128,256,512]  & 5                 & 2     & 1       \\
    $Z_1Z_2$model& [256,128,64,32]& [32,64,128,256]  & 4                 & 3        & 1/8    \\
    XXZ model& [8192,4096,...,256,128]& [128,256,...,4096,8192]  & 8  & 100  & 0.1\\
    MG model& [512,256,128,64]& [64,128,256,512]  & 50                 & 50      & 1    \\
    232 model& [512,256,128,64]& [64,128,256,512]  & 50                 & 50      & 1       \\
    \bottomrule
  \end{tabular}\label{Tab:NNsettings}
  \vspace{0.5em}
\end{table*}

\bibliography{VGON}

\end{document}